\documentclass[preprint,review,12pt]{elsarticle}

\usepackage{amssymb} 
\usepackage{amsmath,bm} 
\usepackage[ruled]{algorithm2e} 
\usepackage{colortbl,booktabs} 
\usepackage{tabularx} 
\usepackage{threeparttable} 
\usepackage{multicol,multirow} 
\usepackage{graphicx}
\usepackage{subfigure} 
\usepackage{enumerate}
\usepackage [font=footnotesize,labelfont=bf]{caption} 
\usepackage{rotating} 
\usepackage{tikz} 
\usepackage[colorlinks,linkcolor=blue]{hyperref}

\usepackage{color}
\definecolor{R}{rgb}{0.9, 0, 0}
\definecolor{G}{rgb}{0, 0.9, 0}
\definecolor{B}{rgb}{0, 0, 0.9}

\definecolor{bzhangCol}{rgb}{0.9, 0, 0.2}
\definecolor{czhangCol}{rgb}{0.9, 0, 0.2}
\definecolor{xyhuCol}{rgb}{0.9, 0, 0.9}

\journal{Elsevier}

\begin{document}
	\begin{frontmatter}
		\title{A method for automated regression test in scientific computing 
			libraries: illustration with SPHinXsys}
	    \author{Bo Zhang}
	    \ead{bo.zhang.aer@tum.de}
	    \author{Chi Zhang}
	    \ead{c.zhang@tum.de}
	    \author{Xiangyu Hu\corref{mycorrespondingauthor}}
	    \cortext[mycorrespondingauthor]{Corresponding author.}
	    \ead{xiangyu.hu@tum.de}
	    \address{TUM School of Engineering and Design, Technical University of Munich,\\
	    	 85748 Garching, Germany}
	    
	   \begin{abstract}
	   	Scientific computing libraries, either being in-house or open-source, 
	   	have experienced enormous progress in both engineering and scientific research. 
	   	It is therefore essential to ensure that the modifications in 
	   	the source code aroused by bug fixing or new feature development wouldn't 
	   	compromise the accuracy and functionality that has already been validated and verified.
	   	With this in mind, this paper introduces a method for developing and 
	   	implementing an automatic regression test environment and takes 
	   	the open-source multi-physics library SPHinXsys 
	   	\cite{zhang2021sphinxsys} as an example.
	   	Firstly, the reference database for each benchmark test is generated 
	   	from monitored data by multiple executions.
	   	This database contains the maximum variation range of metrics for different 
	   	types of strategies, i.e., time-averaged method, ensemble-averaged method 
	   	as well as the dynamic time warping method, covering the uncertainty 
	   	arising from parallel computing, particle relaxation, physical instabilities, etc.
	   	Then, new results obtained after source code modification will be tested 
	   	with them according to a curve-similarity based comparison.
	   	Whenever the source code is updated, the regression test will be 
	   	carried out automatically for all test cases 
	   	and used to report the validity of the current results. 
	   	This regression test environment has already been implemented in all 
	   	dynamics test cases released in SPHinXsys, 
	   	including fluid dynamics, solid mechanics, fluid-structure interaction, 
	   	thermal and mass diffusion, reaction-diffusion, and their multi-physics coupling, 
	   	and shows good capability for testing various problems. 
	   	It's worth noting that while the present test environment is built and 
	   	implemented for a specific scientific computing library,
	   	its underlying principle is generic and can be applied to many others.
  	   \end{abstract}
    
\end{frontmatter}

%
%
\section{Introduction}\label{introduction}
The development of computers has pushed scientific computing to become 
an indispensable part of many technologies and industries, such as in 
assessing climate change \cite{drake2005overview},
designing new energy conductors \cite{post2004software}, 
and imposes an ever-widening
effect on better predicting and understanding the phenomena of 
nature and engineered systems.
Following the definition of validation and verification of scientific 
computing claimed by William \cite{Oberkampf2010}, scientific computing 
should always represent the real world and the conceptual model accurately.
However, this is a challenging task due to the complex mathematical models 
and calculations, which often require changes to separate parts of 
the application and definitely increase the possibility of errors.
Moreover, the development of scientific applications is a lasting work, 
and changes occur frequently due to different requirements 
and new features introduced.
The applications should always produce trustworthy results   
and assure the application qualities 
with the help of validation and verification 
conducted alongside both application development and 
usage \cite{farrell2011automated}.

Implementing testing, including unit test, integration test, regression test, 
system test, etc., can provide concrete validation and verification procedures.
Notwithstanding the fact that it has already been adopted in many IT software, 
scientific applications, especially some open-source libraries, find it difficult 
to perform those testing directly with traditional techniques.
One main challenge is due to the characteristics of the scientific application, 
as it's hard to find a test oracle to check if the program can gain the expected 
output when executing test cases \cite{kanewala2014testing,remmel2011system}.
Another challenge arises from cultural differences between scientists and 
the software engineering community \cite{kanewala2014testing}.
Many scientific applications are usually developed by small-group scientists, 
who may not be very familiar with accepted software engineering practices, 
and haven't studied the developing process of their software in much detail 
\cite{bojana2018framework}, and therefore may overlook the impact of changes.

Regression test stands a great chance to ensure the output validity of 
scientific applications under development.
It is a re-testing activity, which refers to executing the test suite with 
given inputs and comparing the output with previously stored reference 
results when modifications occur or new features are added.
In this way, developers can make sure that their changes don’t cause any unexpected 
side effects and previous functionalities are still verified \cite{remmel2011system}.
Since taking the regression test for all test cases is generally regarded 
as time-consuming and tedious for large-scale software, 
different automatic regression test techniques 
have already been successfully developed and 
implemented to alleviate this drawback, 
including selection \cite{ural2013regression,agrawal2020effective}, 
minimization \cite{di2015coverage,prasad2012regression}, 
prioritization \cite{harikarthik2019optimal,tahat2012regression}, 
or optimization of test cases in the test suite.  

Concerning the implementation of the regression test in scientific computing 
libraries, different focuses are of interest, for example, rigorous validation and 
verification are taken into account to be confident with the computational result.
Therefore, different strategies should be followed and those processes 
should be automated and continuous \cite{farrell2011automated}.
Lin et al. \cite{lin2018exploratory, lin2018hierarchical} have 
adopted historical data of multiple inputs and their relationships 
to define test oracle and then conduct the metamorphic test.
Peng et al. \cite{peng2020unit, peng2021unit} reported their analysis regarding
released unit tests and regression tests of SWMM, a stormwater management model 
developed by the U.S. Environmental Protection Agency. They focused on the test 
coverage and reveal an immature new pattern to mitigate the oracle problems.
Farrel et al. \cite{farrell2011automated} built an automated verification test 
environment for Fluidity-ICOM, which is an adaptive-mesh fluid dynamics simulation package.  
An web-based automated testing environment has also been developed by 
Liu et al. \cite{liu2009automated} to conduct the validating test 
for their computational fluid dynamics (CFD) cases 
related to high-speed aero-propulsive flows. 
Happ \cite{happlinux} developed a set of Linux C-Shell testing 
scripts for SHAMRC, a 2D and 3D finite-difference CFD code solving airblast-related 
problems, and run the regression test daily.

Despite the above-mentioned attempts of introducing testing in mesh-based 
scientific libraries, building a testing environment for meshless libraries 
is still absent even though more and more meshless computing libraries  
have been released and obtained more and more attraction. 
Actually, we encountered several instances when updating the source code of the 
SPHinXsys, an open-source multi-physics library developed by our group,
that one or several test cases that passed the CTest(CMake Test)\cite{KitwareCmake}, 
but unexpectedly lead to the crash of simulations without any error output.
We have believed that such cases were correct as they passed the tests,
but in reality, they fall into fault.
This issue is troublesome and 
has become the main motivation to build an effective regression test environment.

In this paper, a methodological framework for building an automatic 
regression test environment is introduced. 
Firstly, the verified reference database for each test case is generated 
according to adopting different types of monitoring data and stored as the reference. 
Then, the results obtained by the new version code will be tested with the reference database 
according to a curve-similarity based comparison.
Updating the source code will activate the regression test automatically 
for each test case and report the validation of the result.
Such a regression test scheme has already been implemented for all 
test cases released in SPHinXsys, covering different features.
To the best knowledge of the authors, 
there is pioneering work in the regression test for open-source 
scientific computing libraries based on the meshless method. 
This work may also draw some attention from 
the general scientific computing communities to emphasize testing, 
because if the software is meant to do something, then that can and should 
be tested \cite{baxter2004software}.
It's worth noting that the principle presented in this work is versatile
and can also be adopted in other scientific computing libraries.
In what follows, Section \ref{background} provides information about SPHinXsys 
and how tested data is obtained; Section \ref{method} shows an overview of the 
regression test procedure as well as the detailed algorithm of three testing strategies;  
Section \ref{environment} introduces the environment set up and Section \ref{example}
illustrates several implementation examples in SPHinXsys. 
Finally, Section \ref{conclusion} concludes and puts forward future works. 
%
%
\section{Background}\label{background}
In this section, the feature of the SPHinXsys library is briefly introduced, 
and then the method for obtaining different tested data is also explained.

\subsection{SPHinXsys}\label{subsec:SPHinXsys}
As a fully Lagrangian meshless method, smoothed particle hydrodynamics (SPH) was 
originally proposed for astrophysical applications \cite{lucy1977numerical,gingold1977smoothed}, 
and has been applied in simulating a great variety of scientific problems.
SPHinXsys is an open-source multi-physics and 
multi-resolution scientific computing library \cite{zhang2021sphinxsys} based on SPH, 
aiming at solving complex industrial and scientific applications.
The current released version has several important features, such as dual-criteria 
time-stepping, spatio-temporal discretization, multi-resolution, position-based 
Verlet time-stepping scheme, etc., and could efficiently model and solve complex 
systems including fluid dynamics \cite{zhang2017weakly,zhang2020dual}, 
solid mechanics \cite{zhang2021simple}, 
fluid-solid interaction (FSI) \cite{zhang2021multi}, 
thermal and mass diffusion \cite{zhang2021integrative}, 
reaction-diffusion \cite{zhang2021integrative}, 
and electromechanics \cite{zhang2021integrative}, etc.
For quantitative validation, it contains more than 80 test cases 
where the analytical solution, experimental data, or numerical results 
from the literature are available for comparison.
Note that several other open-source scientific libraries 
based on SPH have also been developed 
and released for public users. They all make valuable contributions to the SPH 
community, and their features have been described in the literature, such as 
GPUSPH \cite{bilotta2016gpusph}, SPHysics \cite{GOMEZGESTEIRA2012289}, 
DualSPHysics \cite{crespo2015dualsphysics}, AQUAgpushp \cite{cercos2015aquagpusph}, 
GADGET-2 \cite{springel2005cosmological}, and GIZMO \cite{hopkins2017new}. 
The majority of those open-source scientific applications/libraries, including 
SPHinXsys, are still under intensive development.
Therefore, it is essential to introduce a regression test environment 
for consistent development and release.

\subsection{Obtaining tested data}\label{subsec:tested data}
For the majority of scientific computing problems, 
it is uncommon to obtain a database for the computational domain, 
since monitoring variables of interest
at some typical locations are already able to provide a good reference. 
A variable of interest is usually called a variation point
and its potential values in different executions are called variants \cite{pohl2005software}.
As an example, in CFD simulations, the pressure probed at a 
fixed position can be a variation point and its values at a specific physical 
time for different executions are variants.
By combining the variation point and variants with their constraints, 
a variability model can be generated from a series of computing results 
and provide references for the regression test.

In SPHinXsys, the monitoring data, also known as the variation point, 
can be classified into two types.
One is the observed quantity at probes located within the computational domain.
This includes variables such as density, pressure, velocity, etc., in fluid
problems, and deformation, stress, displacement, etc., in solid,
and other representative variables of interest. 
Another is the reduced quantity, which represents 
the overall variables of interest in the computational domain, such as summation, maximum, 
and minimum values of a variable, e.g., the total mechanical energy of the field. 
Note that these two types of quantities not only serve as the data source 
for the regression test but also for visualization.
\section{Methodology}\label{method}
In this section, the overview of the regression test procedure will be introduced first,
and then the individual steps and algorithms for three testing strategies are detailed follow.

\subsection{Overview of the regression test}\label{subsec:method overview}
The underlying principle of the regression test is to compare the similarity 
between the verified curves (or time series results) generated from the 
previously verified executions in the reference database and the newly 
obtained ones after code modification.
A verified curve usually includes a tolerance
range due to the uncertainty induced in executions.
For example, the concurrent vector is widely used in the shared memory parallel 
programming library, such as the Threading Building Blocks (TBB) library, to construct 
a sequence container with the feature of being concurrently grown and accessed.
However, the results of multiple executions with the same model may not be  
exactly the same, but with noticeable or even considerable differences, 
especially for highly non-linear problems such as fluid dynamics.

\begin{figure}[htbp]
	\centering
	\makebox[\textwidth][c]{\includegraphics[width=1.2\textwidth]{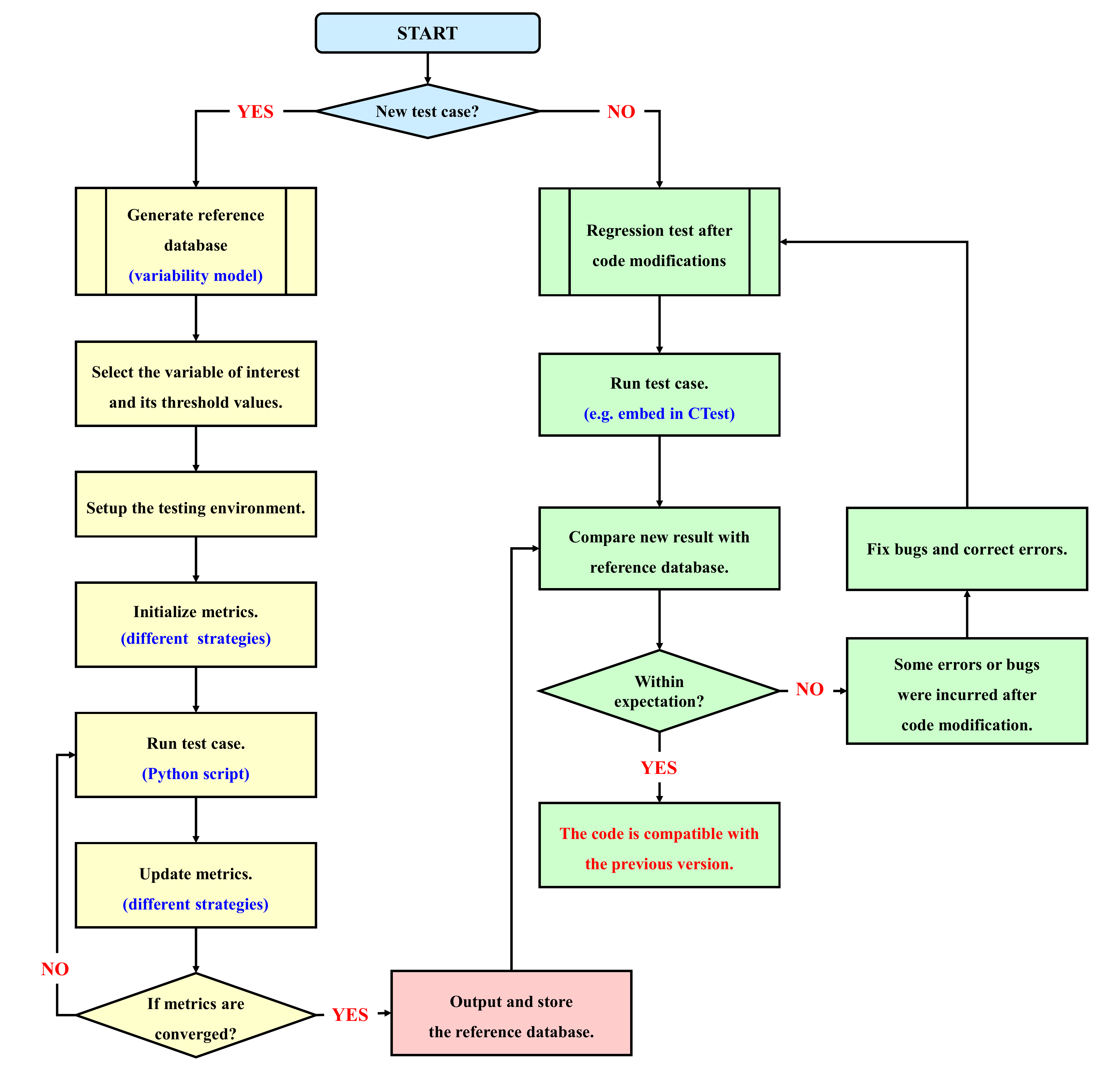}}
	\caption{The flow chart for the regression test. 
		     The left part is for generating the reference database, 
	 	     and the right is for testing the new result obtained after code modification.}
	\label{flowchart}	
\end{figure}

In general, the regression test,
as shown by the flow chart in Fig. \ref{flowchart},
includes two parts:
1) a reference database for each test case is generated; 
2) the new result after code modifications is checked automatically 
following specific strategies.

For a newly added test case, the following steps can be followed to 
generate a reference database:
\begin{enumerate}[]
	\item Step 1: Execute the test case and verify the current result with experimental,
	numerical, or analytical data from the literature.
	\item Step 2: Select one or more variables of interest and define the 
	corresponding thresholds.
	\item Step 3: Set up the testing environment by instantiating objects 
	and introducing methods for the regression test, etc.
	\item Step 4: Choose and initialize metrics from different strategies according 
	to the type of generated curves or monitored time series results.
	Three different strategies, as explained in the following sections
	\ref{subsec:time averaged} - \ref{subsec:DTW}, 
	are applied in this work.
	\item Step 5: Execute the test case multiple times
	and update metrics with different strategies.
	\item Step 6: Until the variations of all metrics are converged under the given threshold, 
	the reference database will be stored. 
\end{enumerate}

Having the reference database in hand, the modified code can be tested as following steps:
\begin{enumerate}[]
	\item Step 1: Run the test case by using the CMake Test or other 
	similar testing packages.
	\item Step 2: Compare the newly obtained result with the previously reserved one 
	in the reference database based on curve-similarity measures.
	\item Step 3: Check whether the similarity measure is within the given threshold.
	If it is, the modified code is considered to be acceptable. 
	Otherwise, the source code should be checked for bugs.
	Then, the testing will be conducted again until the similarity measure is considered as acceptable.
\end{enumerate}

\subsection{Curve classification and testing strategies}\label{subsec:curve type}
Different strategies should be adopted for comparing curve similarity 
regarding various types of curves.
For the typical dynamic problems considered in this work,
the curves of time series data can be generally classified into three types,
each corresponding to a different comparison strategy.

The first type, corresponding to the time-averaged strategy, 
represents data series that fluctuate around a constant 
value after reaching steady-state.
\begin{figure}[!htb]
	\centering
	\makebox[\textwidth][c]{\subfigure[]{
			\includegraphics[width=0.5\textwidth]{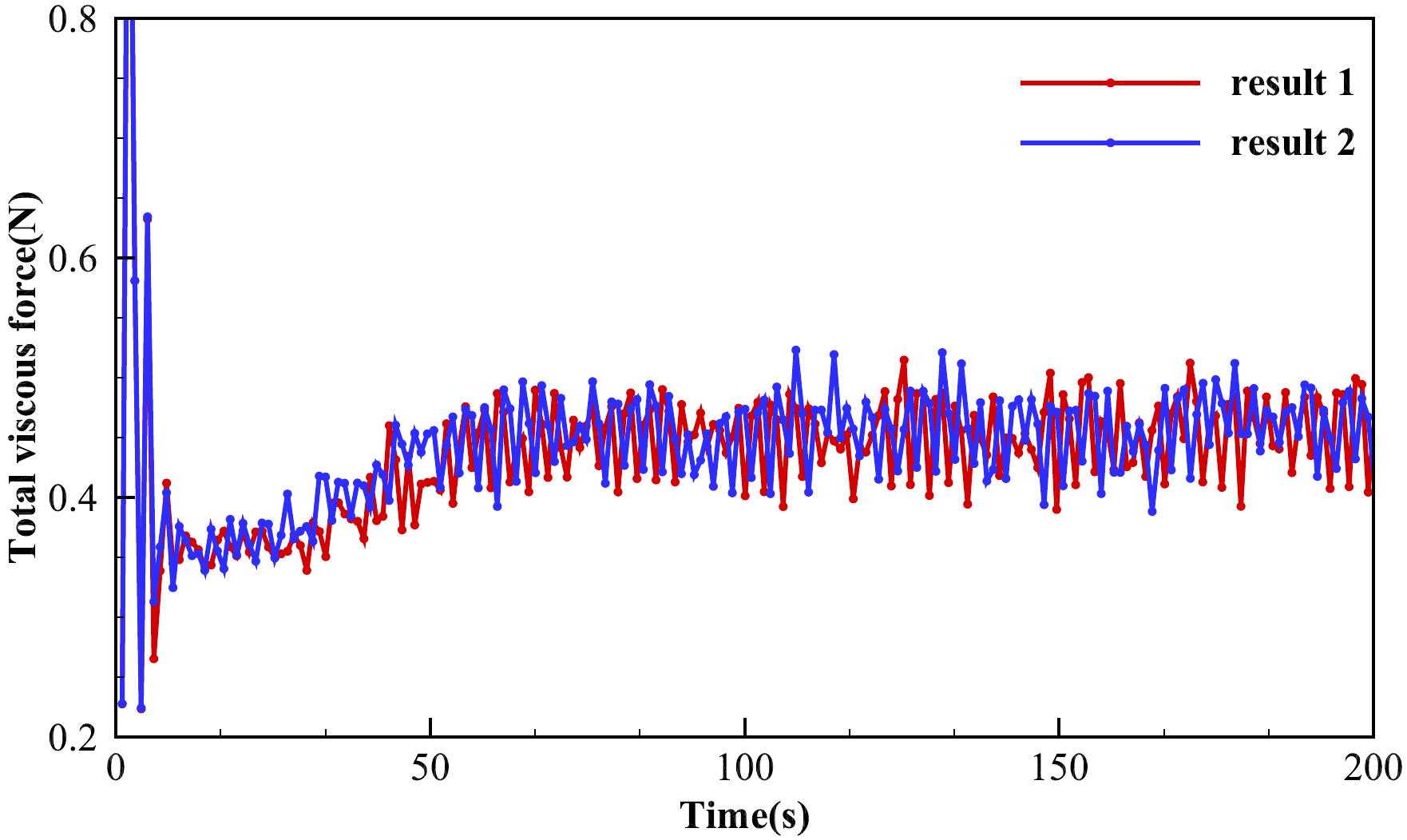}
			\label{FSI-force}
		}
		\quad
		\subfigure[]{
			\includegraphics[width=0.5\textwidth]{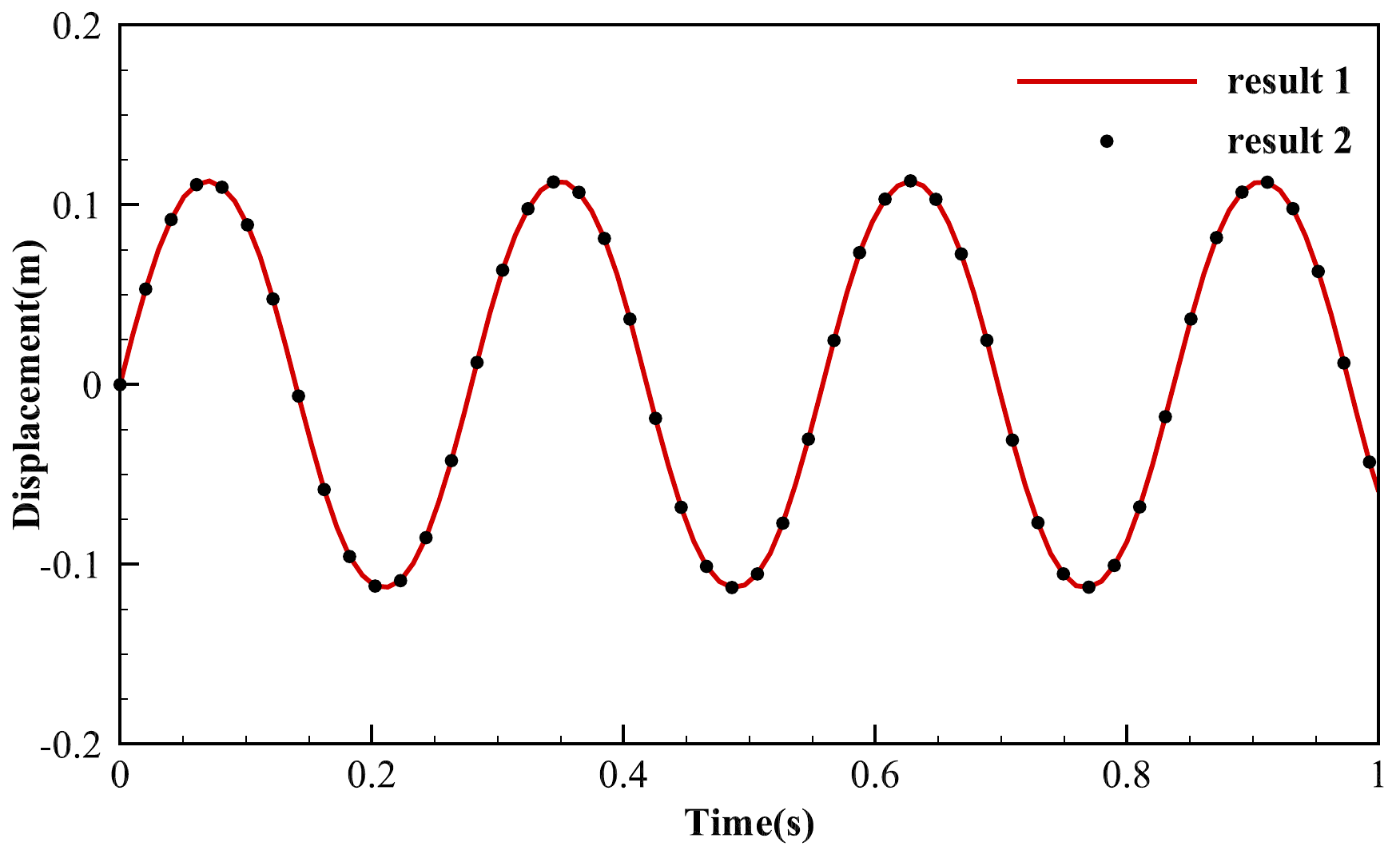}
			\label{oscillating-beam-position}
		}
		\quad}
		\makebox[\textwidth][c]{\subfigure[]{
			\includegraphics[width=0.5\textwidth]{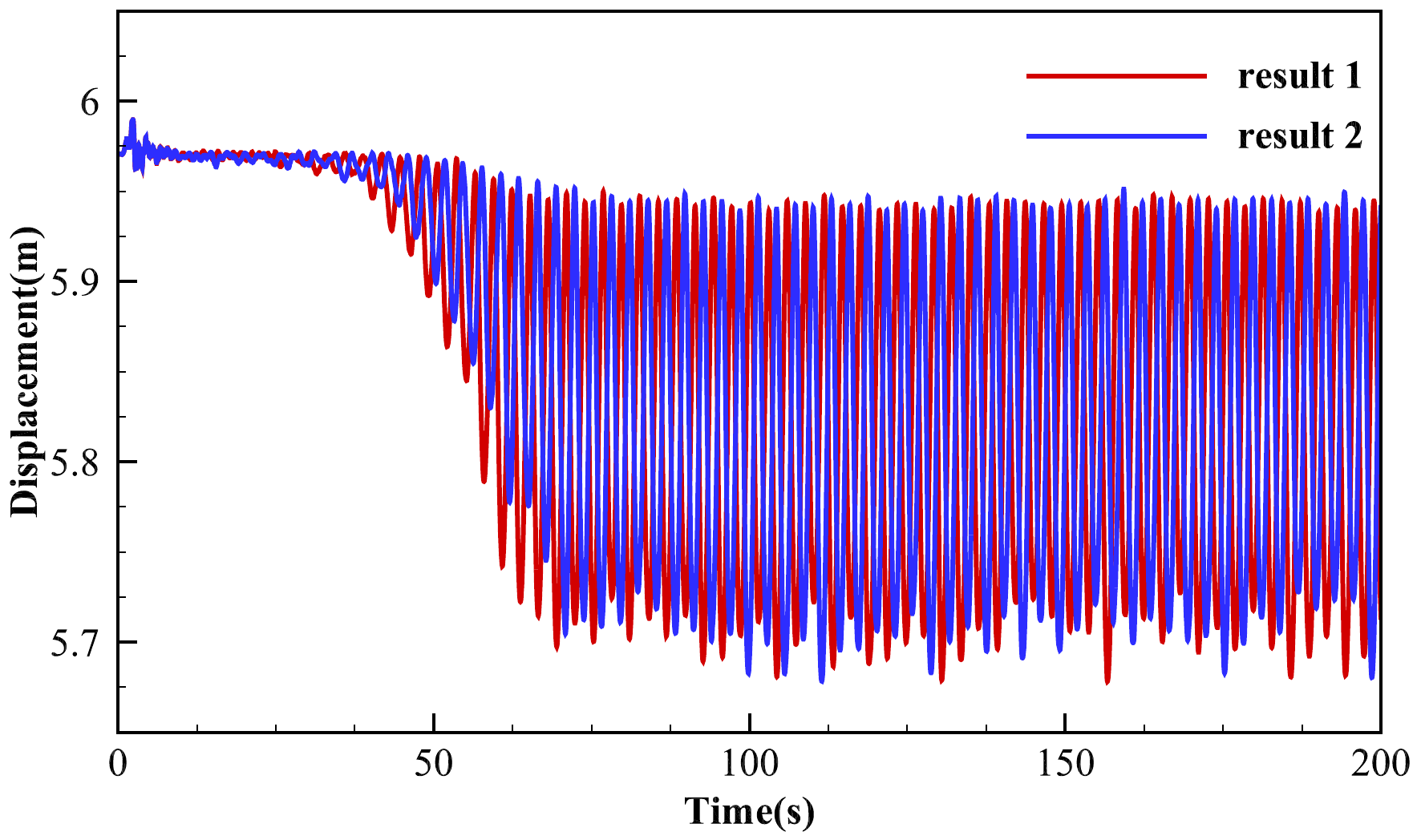}
			\label{FSI-position}
		}
		\quad
		\subfigure[]{
			\includegraphics[width=0.5\textwidth]{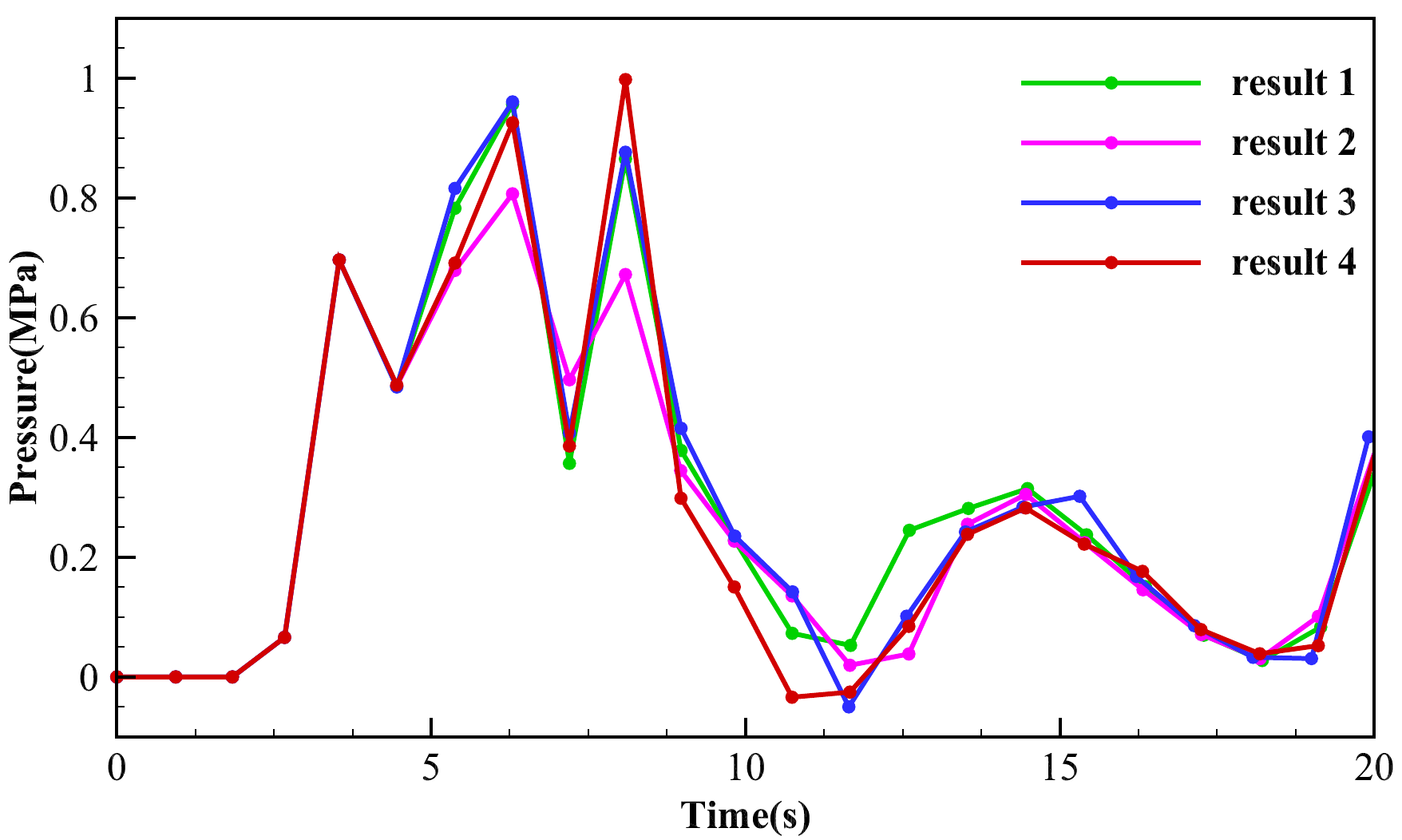}
			\label{pressure}
		}
		\quad}
	\caption{Different types of curves obtained from time series results and used for the regression test.
		(a) Type 1: total viscous force of the flow around cylinder \cite{zhang2021multi}; 
		(b) Type 2: free end displacement of an oscillating elastic beam \cite{zhang2017generalized};
		(c) Type 3: flow-induced free end displacement of an elastic beam attached to a cylinder \cite{zhang2021multi};
		(d) Type 3: solid wall impacting pressure in dambreak flow \cite{zhang2017weakly,zhang2020dual}.}
	\label{figure-type}
\end{figure}
This kind of curve is prevalent in many fluid dynamics problems,
such as the observed total viscous force for an fluid-structure 
interaction (FSI) problem, as illustrated in Figure \ref{FSI-force}.
The second type, corresponding to the ensemble-averaged strategy,
represents data series that exhibit similar variation patterns 
for each computing.
Such curves are often generated from simple solid dynamics problems,
such as the displacement of a given point from the oscillating beam presented in
Figure \ref{oscillating-beam-position}.
The last type, corresponding to the dynamic time warping (DTW) strategy,
represents data series that may experience rapid and scattered variation 
patterns or large high-frequency fluctuation.
Such curves are generally produced in simulations characterized by high nonlinear dynamics.
Figure \ref{FSI-position} displays a monitored position from an FSI simulation, 
and Figure \ref{pressure} shows monitoring pressure for dambreak flow, 
both of which experience obvious variations in each execution.

\subsection{Time-averaged strategy}\label{subsec:time averaged}
In the time-averaged strategy, since the result always enters a steady state 
due to the relaxation process, the time-averaged mean and variance 
are used as metrics for comparison and testing.

\subsubsection{Metrics generation and updating}\label{subsubsec::time averaged update}
The generation of the reference database under this strategy involves
updating the time-averaged mean and variance 
through multiple executions until their variations converge. 
For each updating (e.g., the $n$th execution), 
the mean $M^{n}$ and variance $\sigma^{n}$ 
of the obtained result $x$ from the current execution 
can be calculated respectively as
\begin{equation}
	\begin{cases}
		\displaystyle M^{n} = \frac{1}{l-k}\sum_{i=k} ^l x_{i}^{n}	\\
		\displaystyle \sigma^{n} = \frac{1}{l-k}\sum_{i=k} ^l \left(x_{i}^{n}-M^{n}\right)^2
	\end{cases},
	\label{calculate-time-averaged-metrics}
\end{equation}
where $i$ is the index of a data point, 
$l$ is the total number of data points,
and $k$ is the index of the start point of the steady state.
Then, the mean $M^{*}$ and variance $\sigma^{*}$ 
in the regression test metrics are 
updated based on the results from $n$th computations.
Specifically, $M^{*}$ is updated as
\begin{equation}
	M^{*} = \left(M^{n} + M^{*}\times \left(n-1\right)\right)/n.
	\label{update-new-time-mean}
\end{equation}
Note that, instead of storing all previous mean $n$ times,
the summation of the mean is recursively updated as a 
decaying average of all previous means for more efficiency.
$\sigma^{*}$ is updated as
\begin{equation}
	\sigma^{*} = \max\left(\sigma^{n}, \sigma^{*}\right),
	\label{update-new-time-variance}
\end{equation}
indicating that the variance is always updated to the maximum variation range.
After the relative difference between the newly updated metrics and the previous ones 
is smaller than thresholds in several successive executions(usually 4), 
the $M^{*}$ and $\sigma^{*}$ are stored as the reference database.
It is worth noting that the variation of the metrics of the two successive runs 
being smaller than the threshold is only a hint of convergence, 
so such should happen several times successively to ensure 
a real stable convergence.
Therefore, once such variation is larger than the threshold, the counting 
of the converged successive executions will be reset as zero.
\subsubsection{Start point searching}\label{subsubsec:search starting point}
As the large oscillatory result in the early stage is nonphysical
and can't accurately reflect the real physical state of interest,
this portion of data should be excluded.
Here, a searching technique is proposed 
to locate the start point of the steady state,
ensuring a reliable calculation of mean value and variance.
Specifically, the search begins from the end of the time series as 
the simulation time is always set to be sufficiently long to ensure a steady state.
To achieve this, $n$ pieces with the same time interval are 
sampled from the entire data, and, beginning from the end, two successive
pieces are averaged separately and compared.
\begin{figure}[htb!]
	\centering
	\includegraphics[width=1.0\textwidth]{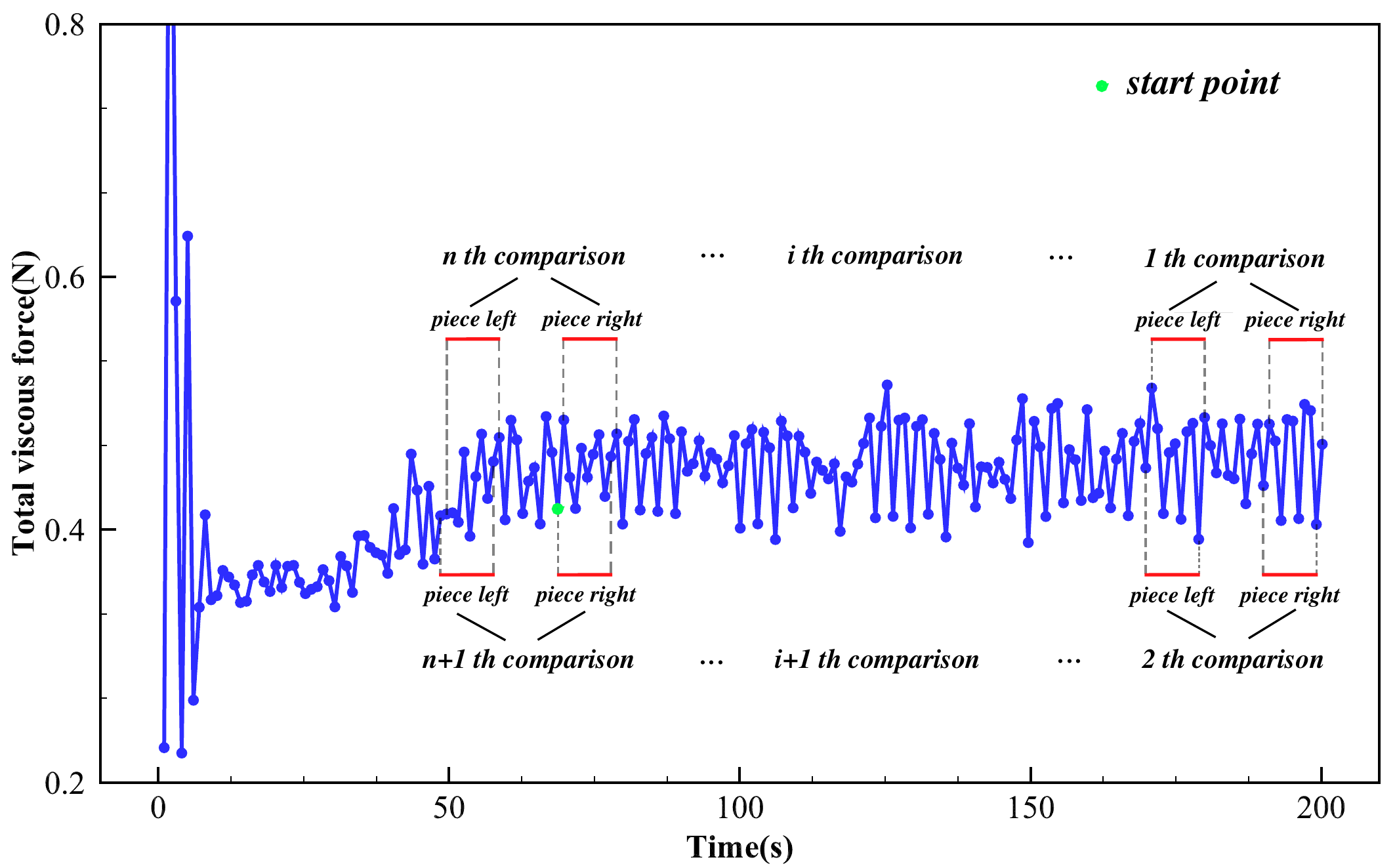}
	\caption{Schematic illustration of the start point searching technique.}
	\label{figure-search-start-point}
\end{figure}
As shown in Fig. \ref{figure-search-start-point}, 
the comparison will proceed until the difference between two averages is larger 
than the given threshold.
\begin{algorithm}[h]
	\caption{: Search start point}
	\label{search-starting-point}
	\LinesNumbered
	\KwIn{The current result $Array$ and its length $l$}
	\KwOut{The index of the start point: $k$}
	Number of data points in each piece: $n = l/20$\;
	\For{$i=L-1$; $i\neq 3 \times n$; $--i$}{
		$M_{1} = 0$, $M_{2} =0$\;
		\For{$j=i$; $j \neq i-n$; $--j$}{
			$M_{1} \mathrel{+}= Array[j]/n$\;
			$M_{2} \mathrel{+}= Array[j-2\times n]/n$\;
		}
		\If{$\lvert \left(M_{1} - M_{2}\right)/\left(M_{1} + M_{2}\right) \rvert \ge \rm{Threshold}$}{
			$k = \max\left(k, i- n\right)$\;
			$break$\;
		}	
	}
\end{algorithm} 
The earliest pieces will be regarded as the start point of steady state.
The detailed procedure is given in Algorithm \ref{search-starting-point}
\subsubsection{Regression test}\label{subsubsec::time averaged test}
For the regression test, the mean value and variance of the new result are 
compared with the metrics of the reference database.
If the following conditions
\begin{equation}
	\begin{cases}
		\lvert M-M^{*} \rvert \le \alpha M^{*}	\\
		\sigma \le \sigma^{*}
	\end{cases}
	\label{time-averaged-regression-test}
\end{equation}
are satisfied, the new result is considered correct
and the modified code is compatible with the previous version.
Here, the parameter $\alpha$ is chosen according to 
different types of dynamics problems, 
and $\alpha=0.05$ for solid dynamics and $\alpha=0.1$ for fluid dynamics
are applied in SPHinXsys.
%
%
\subsection{Ensemble-averaged strategy}\label{subsec:ensemble averaged}
In the ensemble-averaged strategy,
the result curves obtained from the simulation runs 
are often similar to each other with a variation range, 
which is defined by the metrics of ensemble-averaged mean and variance.

\subsubsection{Metrics generation and updating}\label{subsubsec:ensemble averaged update}
For the $n$th execution, 
the metrics for each data point $i$ are updated based on the previous values 
and the new results.
The ensemble-averaged mean $M_{i}^{*}$ at a data point $i$ is updated as
\begin{equation}
	M_{i}^{*} = \left(x_{i}^{n} + M_{i}^{n-1} \times
	\left(n-1\right)\right)/n.
	\label{update-ensemble-mean}
\end{equation}
where $x_{i}^{n}$ is the newly obtained data point 
and $M_{i}^{n-1}$ is the previous mean.
Similar to the time-averaged strategy, 
the new variance $\sigma_{i}^{*}$ is updated as
\begin{equation}
	\sigma_{i}^{*} = \max\left(\sigma^{*}_{i}, \sigma_{i}^{n-1}, \left(0.01*\left(M_{i,max}^{*}-M_{i,min}^{*}\right)\right)^{2}\right),
	\label{update-ensemble-variance}
\end{equation}
where the last term is a secure value that is 
introduced to create a variation range and 
prevent zero maximum variance for results from different computations.
Such a secure value is set according to the maximum and minimum value 
of the local result.
Again, the convergence criteria same as that in the time-averaged strategy,
i.e. successive executions with sufficient small variations of 
the updated mean and variance are used to terminate the metrics updating. 
\subsubsection{Extreme value filter}\label{subsubsec: extreme filter}
In some cases, particularly in certain physical problems,
there will be some obvious extreme values in otherwise smooth and regular results.
These values are often obtained from fast events with insufficient sampling frequency, 
such as wave-impacting events within generally continuous FSI problems,
and may negatively impact the accuracy of the metrics.
Therefore, an extreme value filter is introduced and adopted in some test cases.

As shown in Algorithm \ref{extreme-value-filter},
\begin{algorithm}[h]
	\caption{Extreme value filter}
	\label{extreme-value-filter}
	\LinesNumbered
	\KwIn{The current result $Array$ and its total length $l$}
	\KwOut{The smoothed result $Filtered$}
	Length of each data segment: $n = l/200$\;
	\For{$i=0$; $i\ne l$; $++i$}{
		$Filtered[i] = Array[i], M_{nbh} = 0, \sigma_{nbh}= 0 , \sigma_{test}= 0$\;
		\For{$j = \max\left(i-n, 0\right)$; $j \ne \min\left(i+n,l\right)$; $++j$}
		{
			calculate neighboring points mean $M_{nhb}$;
		}
		\For{$j = \max\left(i-n, 0\right)$; $j \ne \min\left(i+n,l\right)$; $++j$}
		{
			calculate neighboring points variance $\sigma_{nhb}$;
		}
		calculate tested point variance $\sigma_{test} = \left(Array[i]-M_{nbh}\right)^{2}$; \\
		\If{$\sigma_{test} > 4 \times \sigma_{nhb}$}
		{
			$Filtered[i] = M_{nhb}$\;
		}
	}
\end{algorithm}
the total dataset is decomposed into $n$ segments,
and each contains one tested data point 
and its neighboring data points.
Then, for each data segment, 
the standard variance of neighboring data points is compared with 
the variance of between the tested data and the neighboring data. 
If the latter is four times larger than the former, 
this tested data point will be considered an extreme 
value point and is reset to the mean of its neighboring data.
\begin{figure}[htb!]
	\centering
	\makebox[\textwidth][c]{\subfigure[]{
			\includegraphics[width=0.5\textwidth]{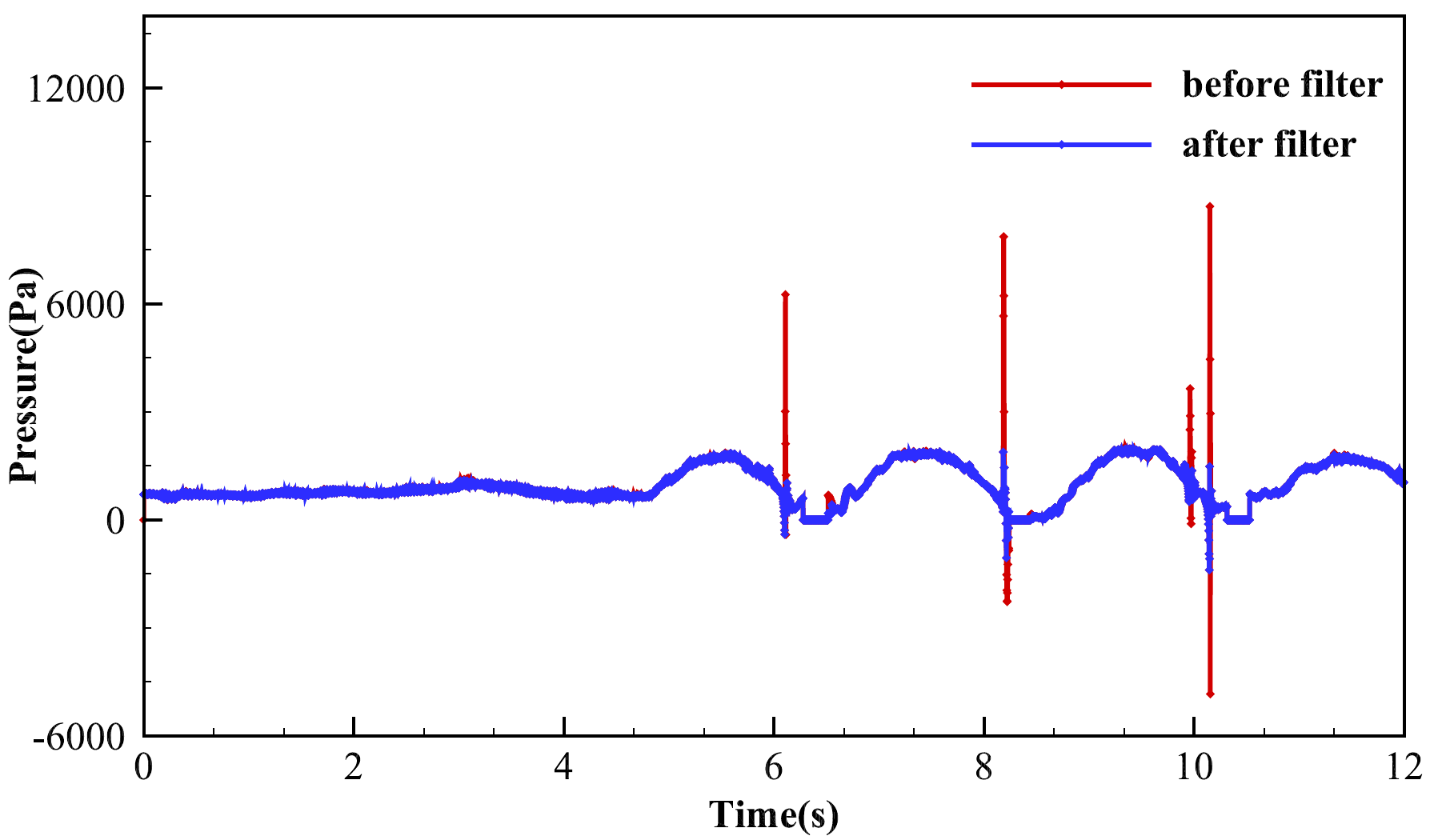}
			\label{pressure-filter}
		}
		\quad
		\subfigure[]{
			\includegraphics[width=0.5\textwidth]{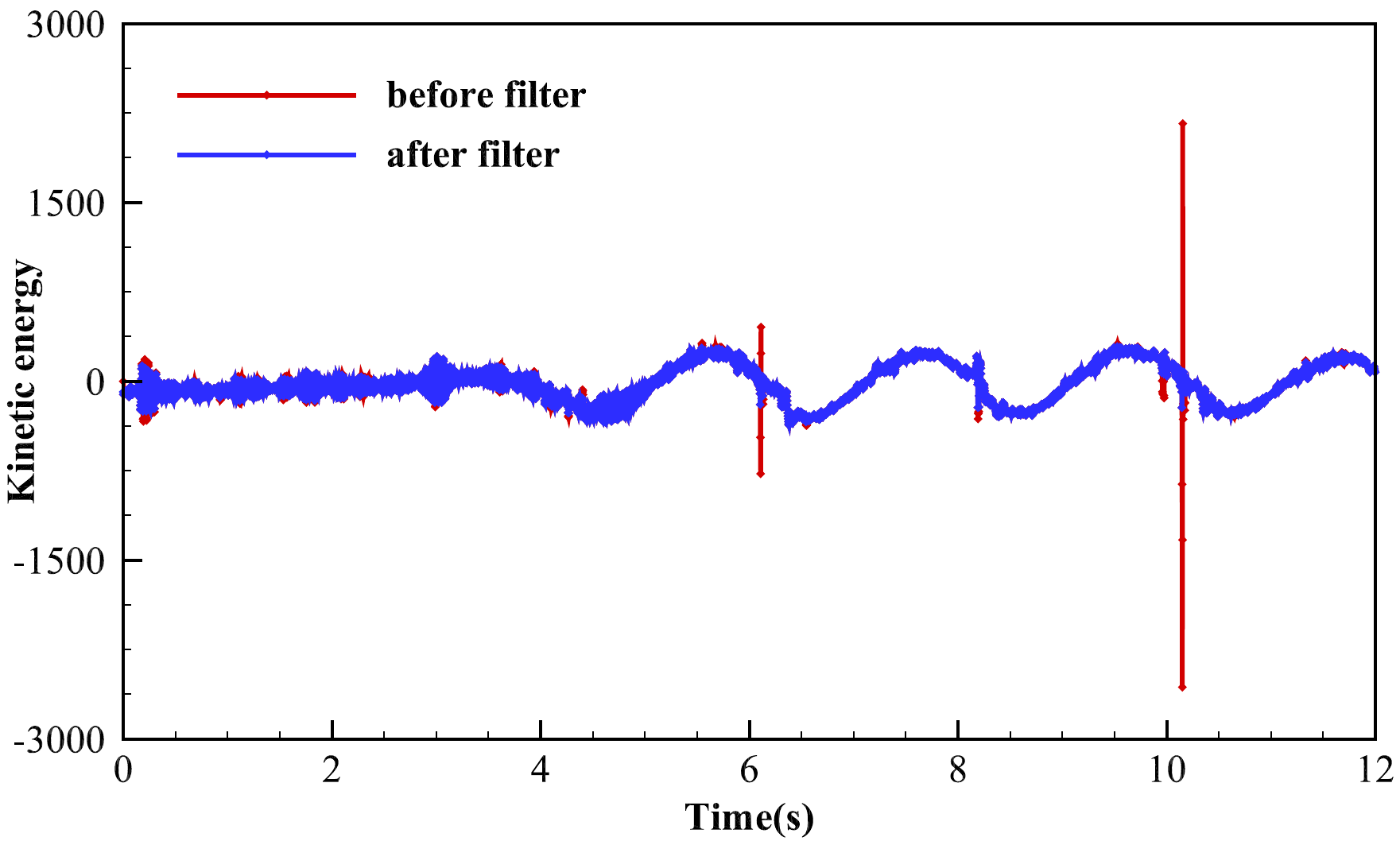}
			\label{energy-energy}
		}
		\quad}
	\caption{Extreme filtering in the problem oscillating
		wave surge converter \cite{zhang2021efficient}: (a) pressure at a probe; 
		(b) total kinetic energy.}
	\label{figure-example-OWSC}
\end{figure}
Fig. \ref{figure-example-OWSC} presents the filtering 
of the pressure of a fixed position 
and total kinetic energy from the oscillating 
wave surge converter case \cite{zhang2021efficient}.
\subsubsection{Regression test}\label{subsubsec:ensemble averaged test}
Having the metrics of the reference database in hand, 
the regression test after code modification 
will be carried out for all data points with the following condition
\begin{equation}
	\displaystyle\sqrt{\left(x_{i}^{n}-M_{i}^{*}\right)^{2}} \le \sigma_{i}^{*}.
	\label{test-ensemble}
\end{equation}
If there is any data point that does not satisfy this condition, 
the code modification should be checked and corrected.

\subsection{Dynamic time warping (DTW) strategy}\label{subsec:DTW}
DTW, which was originally proposed for spoken word recognition 
\cite{sakoe1978dynamic, myers1980performance}, 
is a dynamic programming algorithm used to 
measure the similarity between two sequences 
with temporal variation by computing the DTW distance.
Compared to the Euclidean distance, DTW is more accurate 
and can handle non-linear distortions, shifts, 
and scaling in the time dimension.
It has been widely used in various research areas 
such as sign language recognition \cite{cheng2020chinese} and 
time-series clustering \cite{li2020adaptively}, etc.
Additionally, this algorithm is also applied in many 
engineering fields involving time-series comparison, 
e.g., in health monitoring and fault diagnosis \cite{douglass2018dynamic}.
Actually, due to its generic properties,
the DTW strategy may also be used for the curve types 
that are not classified in Section \ref{subsec:curve type}.
\subsubsection{Calculation of DTW distance}\label{subsubsec:calculate DTW}
\begin{figure}[htb!]
	\centering
	\makebox[\textwidth][c]{\subfigure[]{
			\includegraphics[width=0.5\textwidth]{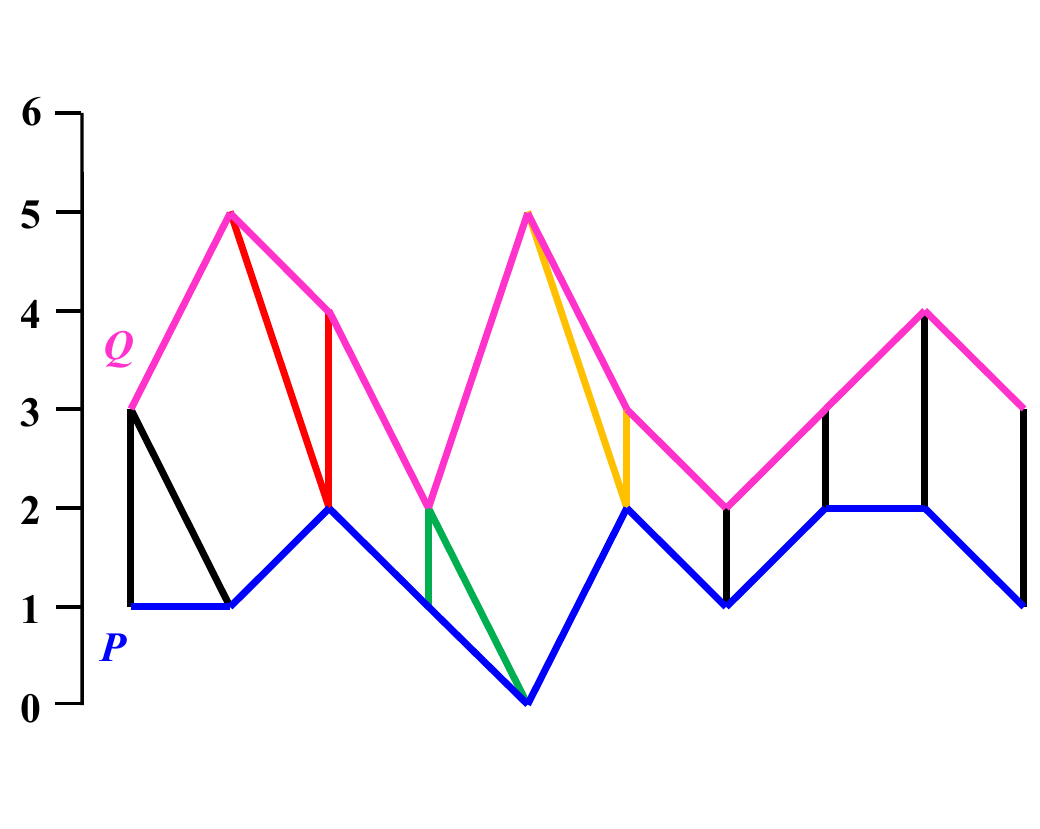}
			\label{DTW-warping-line}
		}
		\quad
		\subfigure[]{
			\includegraphics[width=0.5\textwidth]{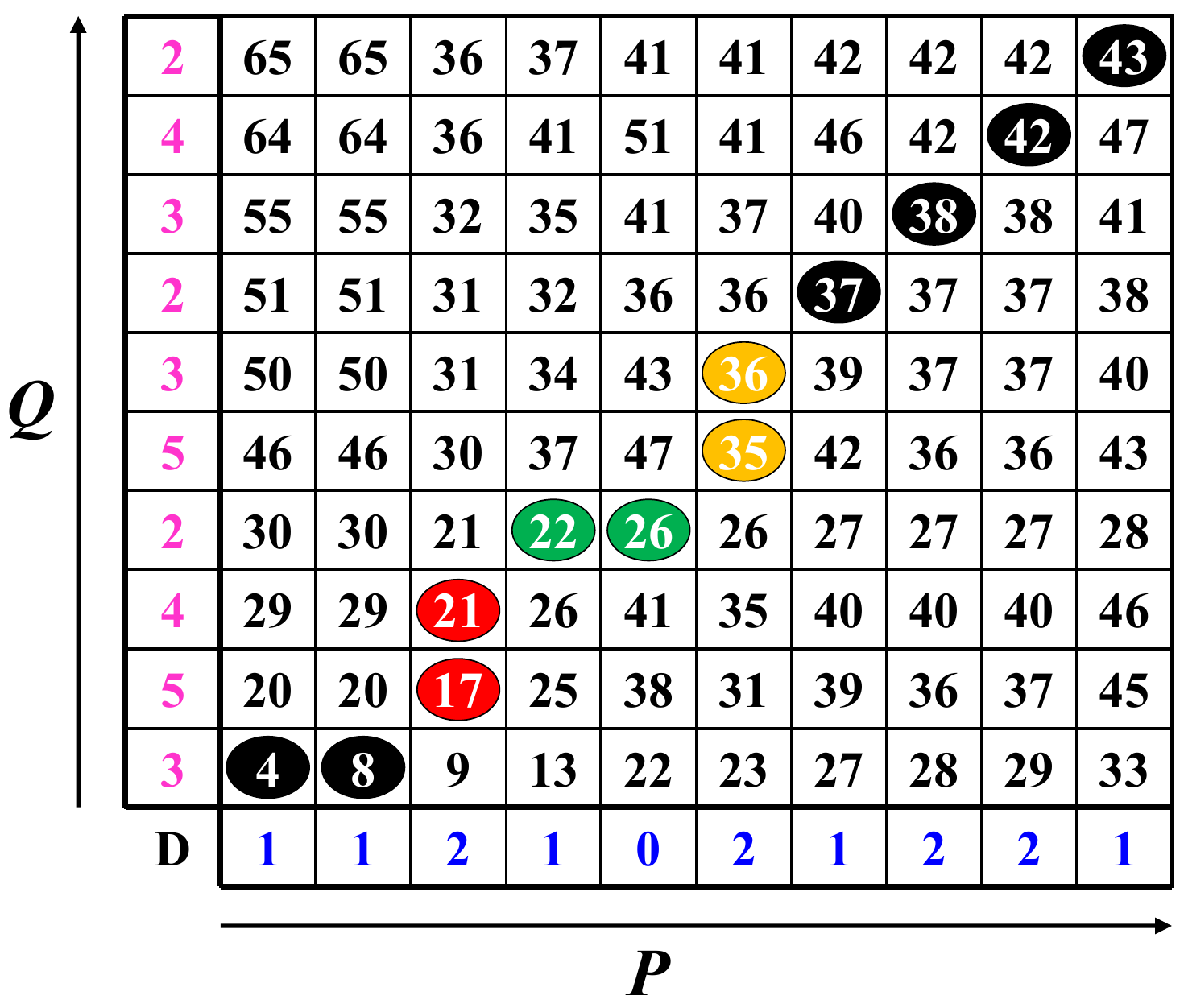}
			\label{DTW-matrix}
		}
		\quad}
	\caption{Illustration of the DTW. 
		(a) Aligning measurements to calculate the DTW distance between two sequences, 
		P and Q;
		(b) The warping matrix D generated by DTW; labeled cells represents the optimal warping path.}
	\label{figure-DTW-calculation}
\end{figure}

Suppose we have two time series, 
\begin{equation}
	\begin{cases}
		P : p_{1}, p_{2}, \dots, p_{i}, \dots, p_{m-1}, p_{m} \\
		Q : q_{1}, q_{2}, \dots, q_{j}, \dots, q_{n-1}, q_{n}
	\end{cases},
	\label{time-series-1}
\end{equation}
where $m$ and $n$ indicate the length of time series $P$ and $Q$, respectively.
$i$ and $j$ are data point indices in the time series.
The DTW algorithm divides the problem into multiple sub-problems, 
and each sub-problem contributes to the cumulative calculation 
of the distance \cite{al2012sparsedtw}.
The first step is to construct a local distance matrix $d$ consisting of 
$m \times n$ elements, where each element represents the Euclidean distance 
between two data points in the time series. 
Then, the warping matrix $D$, seen in Fig. \ref{figure-DTW-calculation}, 
is filled based on 
\begin{equation}
	D(i,j) = d(i,j) +\min
	\begin{cases}
		D(i-1,j)\\
		D(i-1, j-1)\\
		D(i,j-1)
	\end{cases}.
	\label{calcualte-dtw}
\end{equation}
Finally, DTW reports the optimal warping path and the DTW distance.
The warping path consists of a set of adjacent matrix elements 
that identify the mapping between two sequences, representing 
the path that minimizes the overall distance between $P$ and $Q$.
Each warping path should follow certain rules
\cite{sakoe1978dynamic, ratanamahatana2004everything, salvador2007toward}:
each index from the first sequence must be matched with one or more indices 
from the other sequence, and such mapping must be monotonically increasing. 
Note that the first index and the last index from the first sequence must be 
matched with their counterparts from the other sequence correspondingly.

\begin{figure}[htb!]
	\centering
	\includegraphics[width=0.6\textwidth]{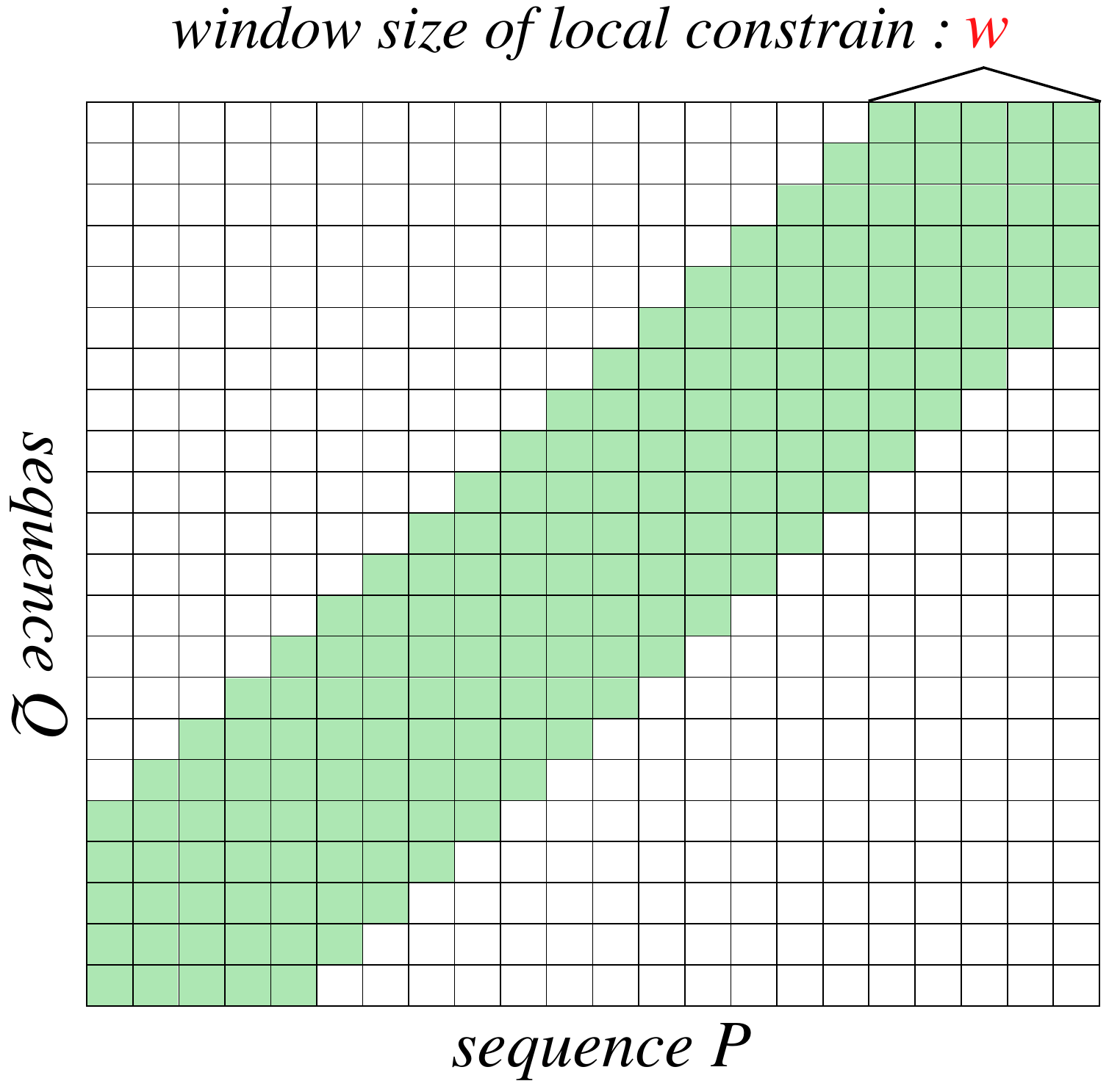}
	\caption{The warping scope limited by the warping window.}
	\label{DTW-scope}
\end{figure}

However, DTW can be computationally expensive when searching for global matches,
and as a result, many algorithms have been proposed to reduce futile computation
\cite{sakoe1978dynamic, salvador2007toward, keogh2005exact, lemire2009faster, 
	sakurai2005ftw}.
One effective and simple method for speeding up DTW is to set 
a warping window(ww)  \cite{sakoe1978dynamic}.
The warping window adds a local constraint that forces the 
warping path to lie within a band around the diagonal,
as shown in Fig. \ref{DTW-scope},
restricting the searching window to the fixed size $w$.
In the current work, we adopt the window size $w$ as
\begin{equation}
	w=\max\left(\lvert m - n \rvert,5\right).
	\label{window}
\end{equation}
After imposing the constraint, the warping will only occur 
within the diagonal green areas, and if the optimal path 
crosses the band, the distance will not be the optimal one.

\subsubsection{Metrics generation and updating}\label{subsubsec:DTW update}
The maximum DTW distance is used as the regression test metric 
and is updated after each execution until
its variation converges to a certain threshold.
With the initial value for the first computation set as $D_{0,0}=0$, 
the maximum distance for the $n$th execution will be calculated as
\begin{equation}
	D^{*}=\max\left(D^{*}, D_{0,n}, D_{1,n}, \dots, D_{n-2, n}, D_{n-1,n}\right),
	\label{new-distance}
\end{equation}
where the subscript, e.g., $D_{n-2,n}$ denote the distance between 
the $(n-2)$th and $n$th computational results.
Similar to the other two strategies,
after the variation of $D^{*}$ converges to a given threshold
in successive several executions, the $D^{*}$ and 
several results (usually 3 $\sim$ 5) with all data points 
are stored for the regression test.
\subsubsection{Regression testing}\label{subsubsec:DTW test}
For the regression test, if the DTW distances between the new result after 
code modification and each result in the reference database are satisfied
\begin{equation}
	\left(D_{1}, D_{2}, \cdots, D_{k}\right) \le D^{*}, k=3\sim5,
	\label{distance-test}
\end{equation}
the new result is regarded as acceptable,
Otherwise, it is beyond expectation 
and the code should be checked and corrected.
%
%
\section{Regression test environment}\label{environment}
\begin{figure}[htb!]
	\centering
	\makebox[\textwidth][c]{\subfigure[]{
		\includegraphics[width=0.5\textwidth]{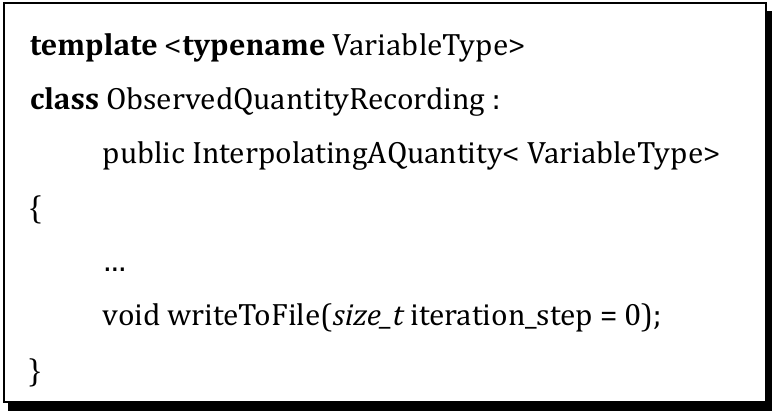}
		\label{observing}
	}
	\quad
	\subfigure[]{
		\includegraphics[width=0.5\textwidth]{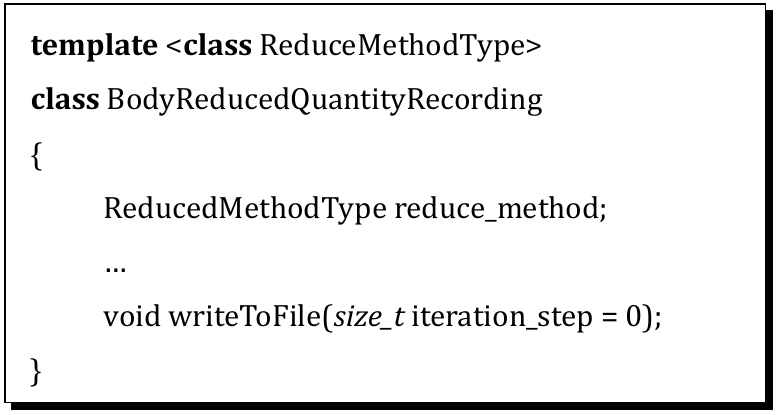}
		\label{reduced}
	}
	\quad}
	\caption{Monitoring methods in SPHinXsys:
		(a) observation of a variable of interest by 
		\texttt{ObservedQuantityRecording}; 
		(b) reducing the variable of interest by 
		\texttt{BodyReducedQuantityRecording}.}
	\label{monitoring-class}
\end{figure}

In this section, taking SPHinXsys as an example, the process of 
building an automatic regression test environment is explained.
The test interface is integrated into a SPHinXsys application's case 
code based on the data monitoring module.
In SPHinXsys, the monitoring module includes two classes,
as depicted in Fig. \ref{monitoring-class}.
The observed quantities at probes are generated from 
\texttt{ObservedQuantityRecording},
and the reduced quantities are obtained from \texttt{BodyReducedQuantityRecording}.
The above two methods are implemented with the template, 
allowing the flexible handling of various data types, and also
providing rich data sources for the regression test.

\begin{figure}[htb!]
	\centering
	\includegraphics[width=0.9\textwidth]{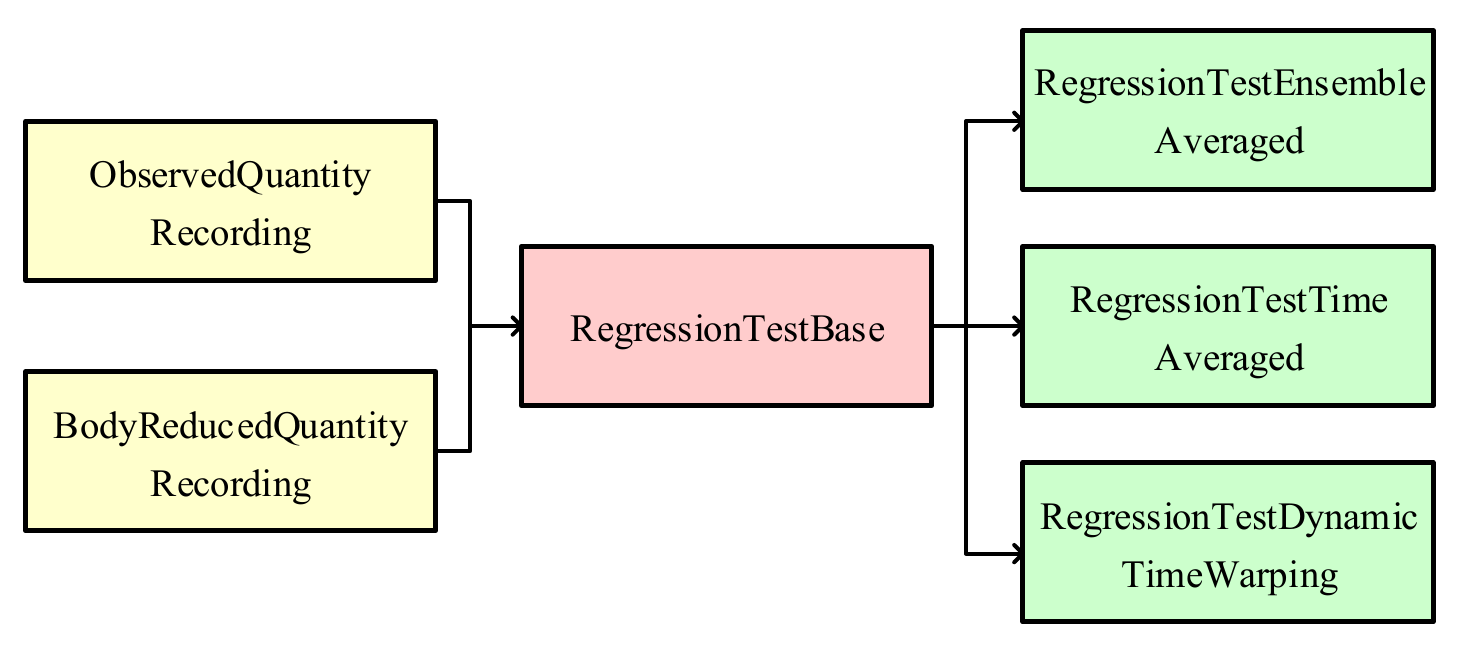}
	\caption{Relationship of classes related to the 
		regression test in SPHinXsys.}
	\label{class-affiliation}
\end{figure}

Fig. \ref{class-affiliation} presents the relationship 
among different regression test methods.
The class template \texttt{RegressionTestBase}, defining commonly used 
methods in the regression test, inherits from the above monitoring class.
Then, three derived template classes are defined to 
implement specialized methods, i.e., 
\texttt{RegressionTestTimeAveraged},
\texttt{RegressionTestEnsembleAveraged} as well as
\texttt{RegressionTestDynamicTimeWarping}, respectively.
Note that the present structure provides a very flexible combination of 
test strategies for various variables of interest.

\begin{figure}[htb!]
	\centering
	\includegraphics[width=1\textwidth]{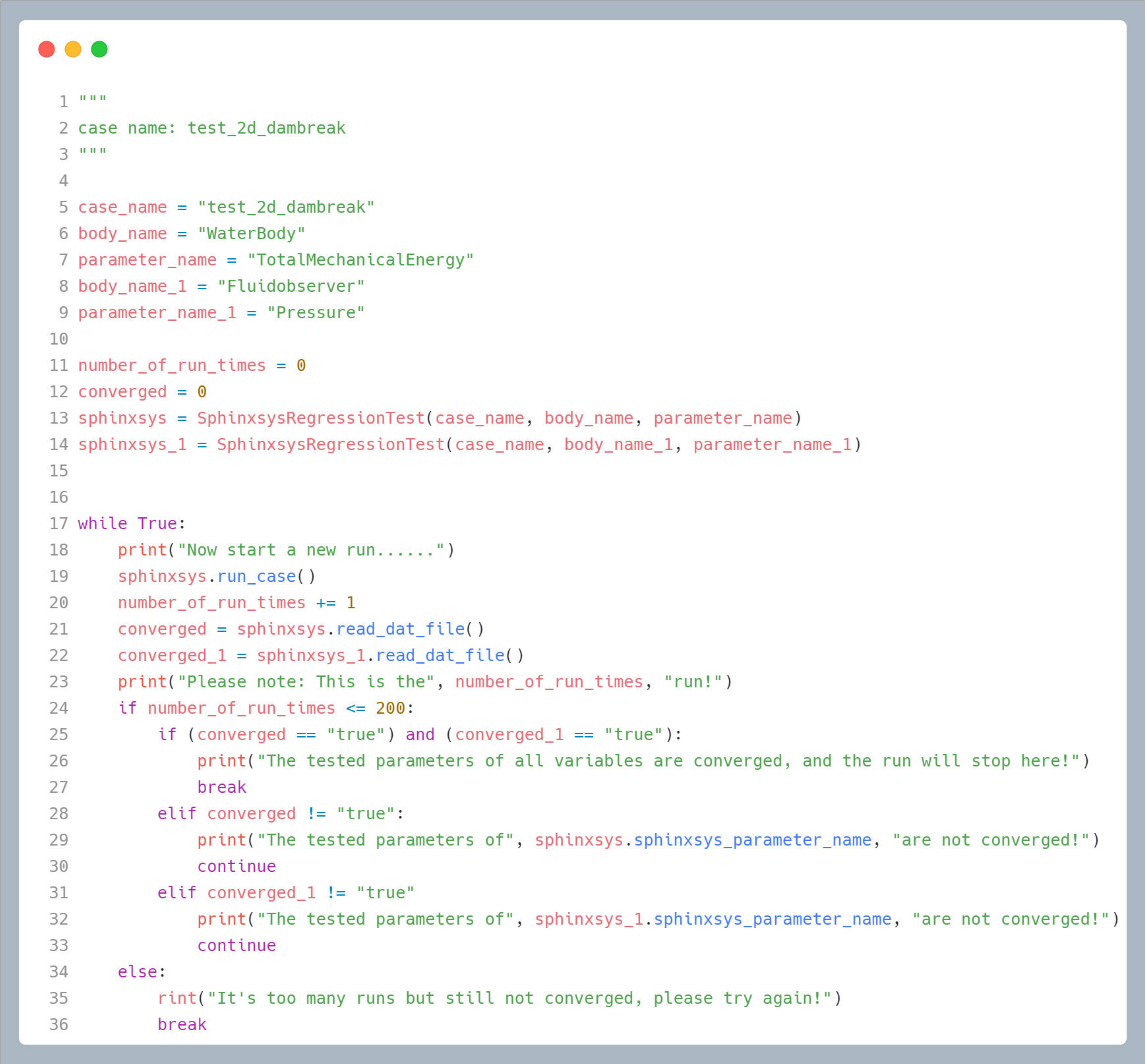}
	\caption{Example of the Python script for generating test metrics
	for a 2d dambreak flow case in SPHinXsys,
    with two variables of interest.}
	\label{python-script}
\end{figure}

To set up a regression test for a specific test case, 
it only needs to replace the existing monitoring class 
with the regression test class based on the type of curve 
that is being monitored. 
Afterward, call the interface \texttt{generateDataBase()} 
to generate a reference database 
or \texttt{testNewResult()} to perform a regression test
at the end of the case file.
It should be noted that the current method does not disturb the existing code structure.
In the SPHinXsys package,
a python script, as exampled in Fig. \ref{python-script},
is employed to execute a test case multiple times 
automatically for generating the reference database.
It is an automatic process as long as the script is made for the test case, 
and it is also easy to regenerate the reference database when it is necessary.

In SPHinXsys, the regression test is setup for all test cases 
which simulate real-world problems with SPH method.
Together with \texttt{Google test} \cite{whittaker2012google} for unit test,
all tests are integrated using the \texttt{CTest} \cite{KitwareCmake}. 
When merging of branches occurs, all tests can be triggered automatically.
%
%
%
\section{Applications and examples}\label{example}
The obtained reference database for several test cases will be presented 
here to demonstrate the functionality of current regression test method.

\subsection{Dambreak}\label{subsubsec:dambreak}
The first example is the dambreak flow in two and three dimensions.
The total mechanical energy of the entire domain
and pressure at fixed probes have been 
recorded and validated \cite{zhang2017weakly, zhang2017generalized}.
Thus, those two monitoring variables have been used for the regression test.
According to the classification of curves, the ensemble-averaged strategy 
is used for the curve of total mechanical energy,
and the DTW strategy is used for the pressure curve.
\begin{figure}[htb!]
	\centering
	\makebox[\textwidth][c]{\subfigure[]{
		\includegraphics[width=0.5\textwidth]{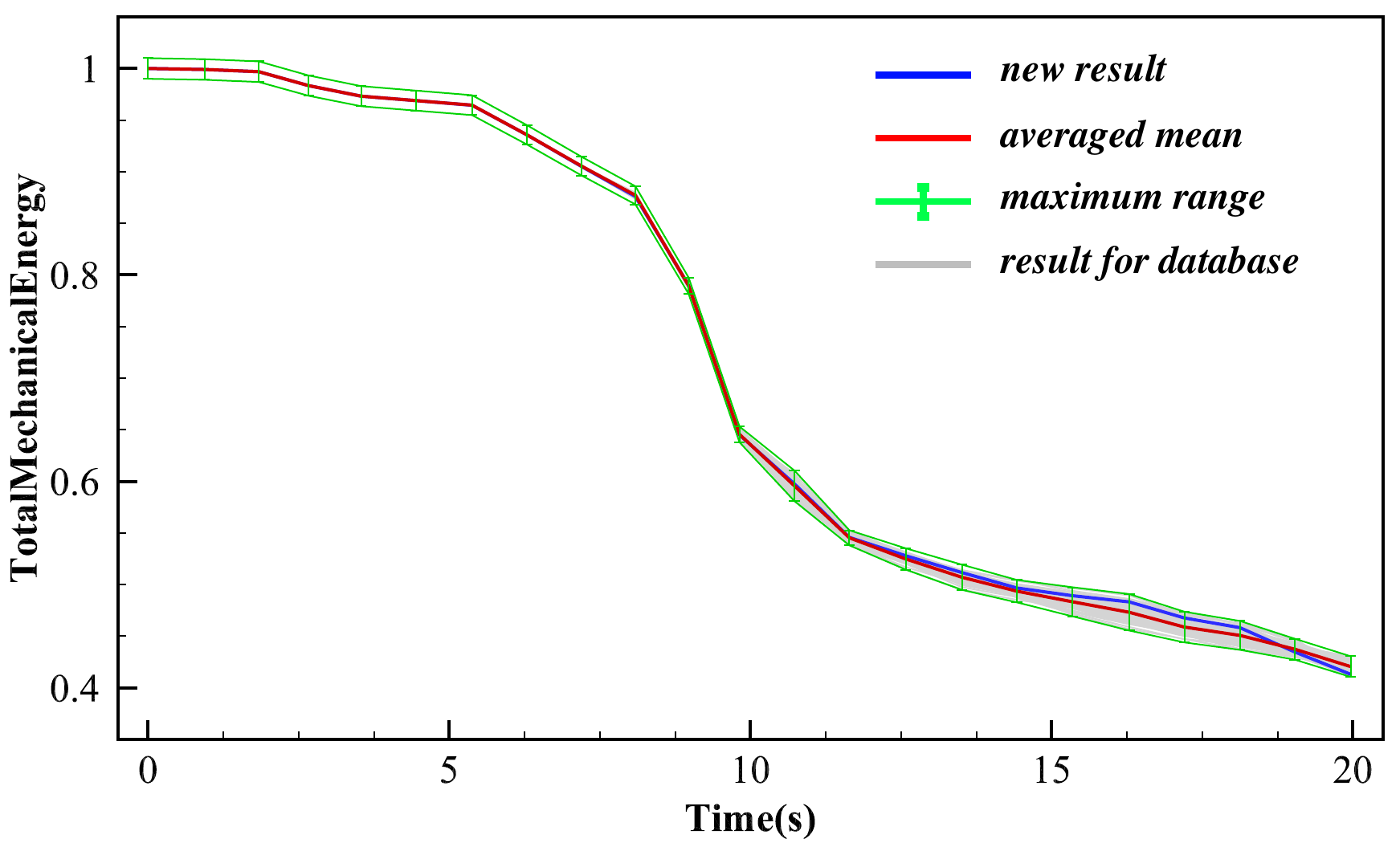}
		\label{me-re-2d}
	}
	\quad
	\subfigure[]{
		\includegraphics[width=0.5\textwidth]{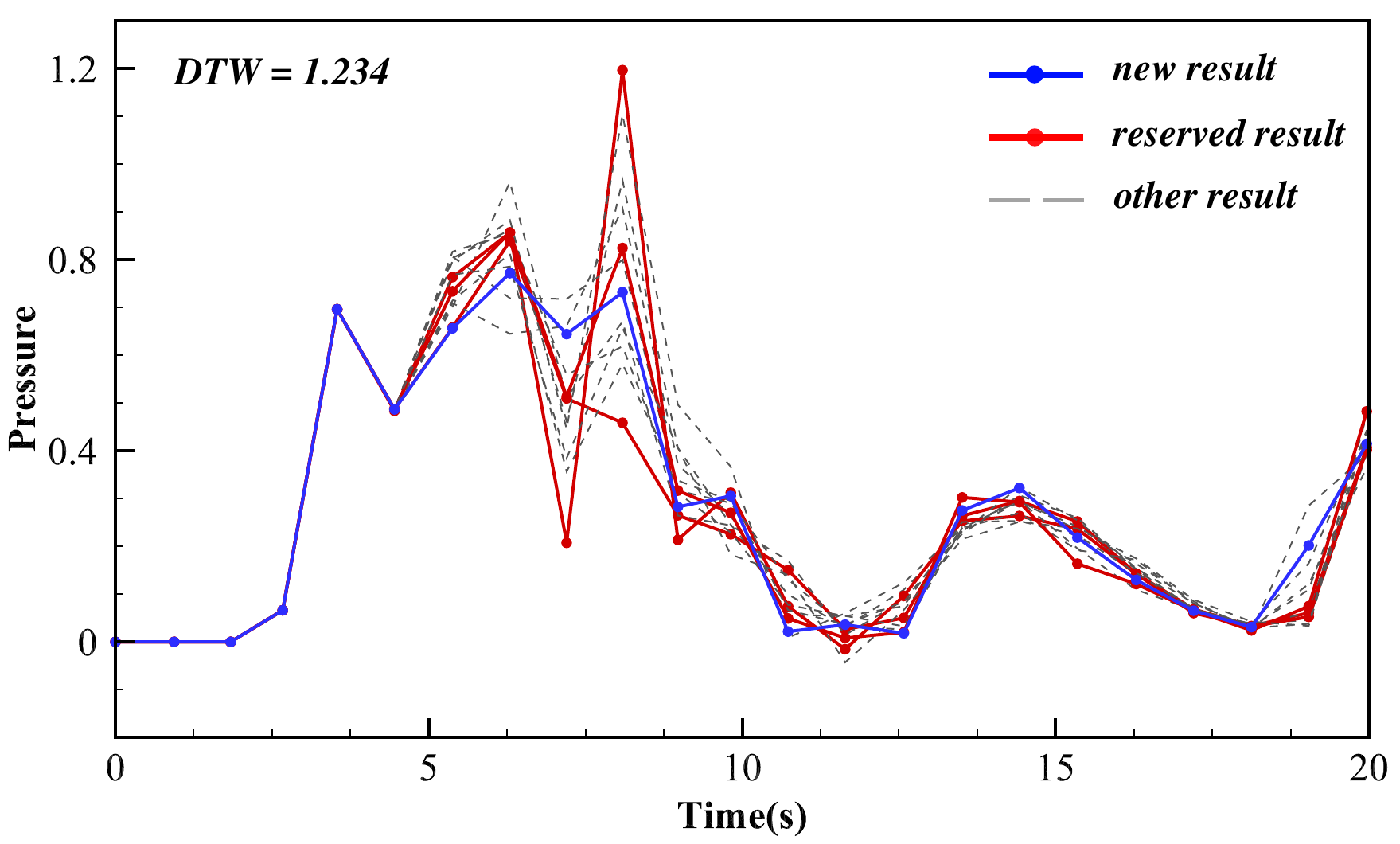}
		\label{pr-re-2d}
	}
	\quad}
	\caption{Illustration of the reference database and results
		     for the 2d dambreak case: (a) total kinetic energy; (b) pressure.}
	\label{dambreak_test_2d}
\end{figure}
\begin{figure}[htb!]
	\centering
	\makebox[\textwidth][c]{\subfigure[]{
			\includegraphics[width=0.5\textwidth]{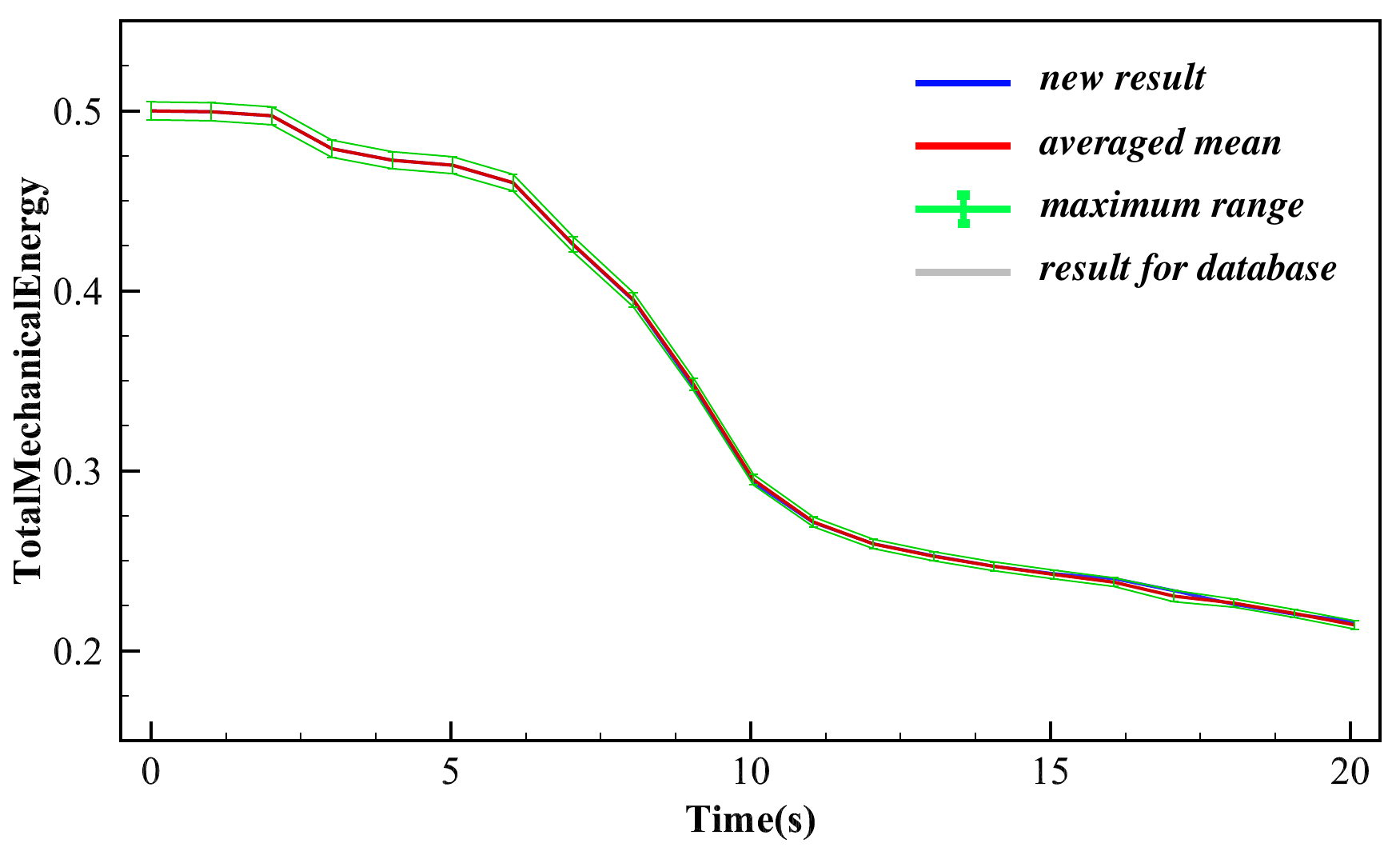}
			\label{me-re-3d}
		}
		\quad
		\subfigure[]{
			\includegraphics[width=0.5\textwidth]{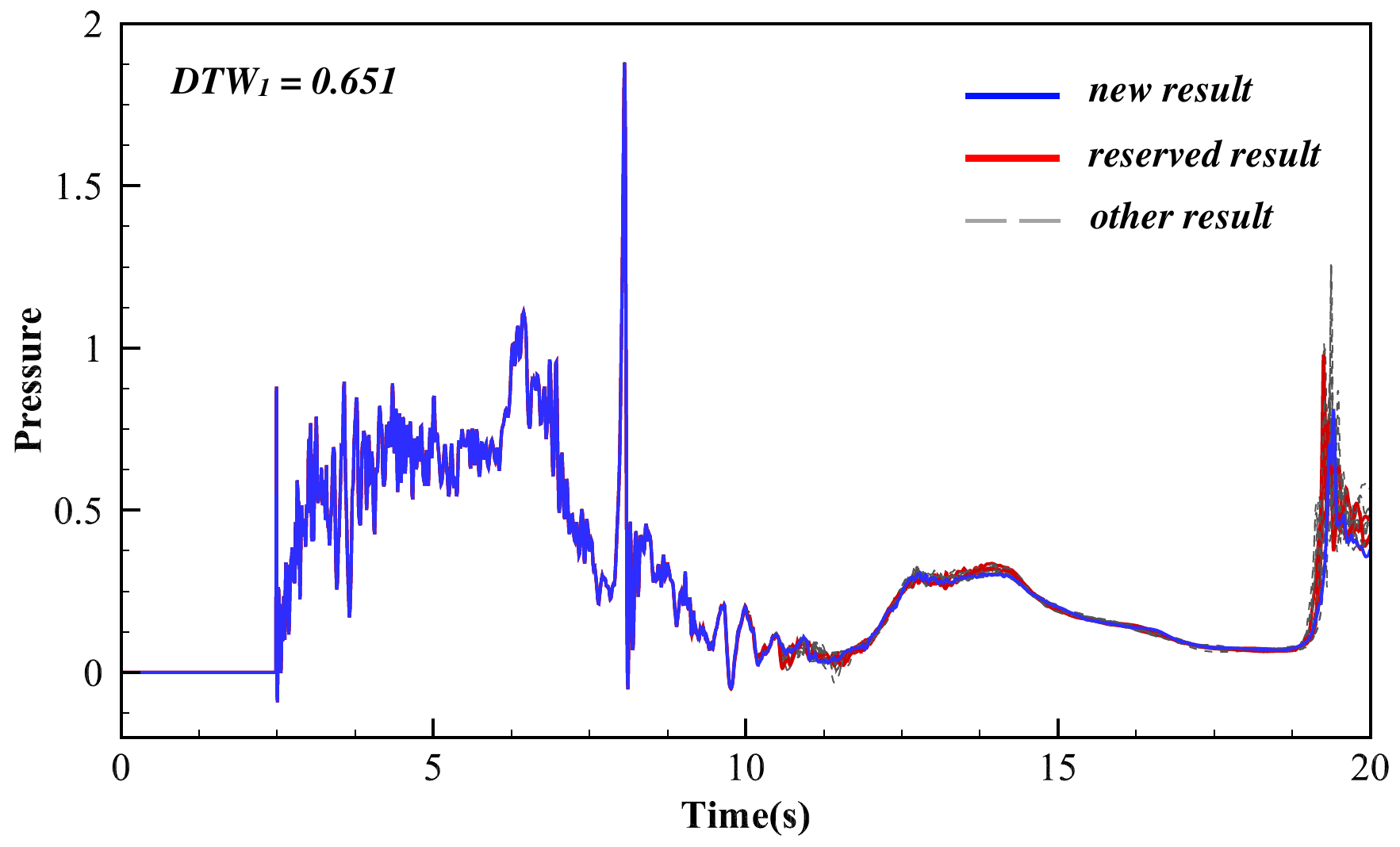}
			\label{pr-re-3da}
		}
		\quad}
	\makebox[\textwidth][c]{\subfigure[]{
			\includegraphics[width=0.5\textwidth]{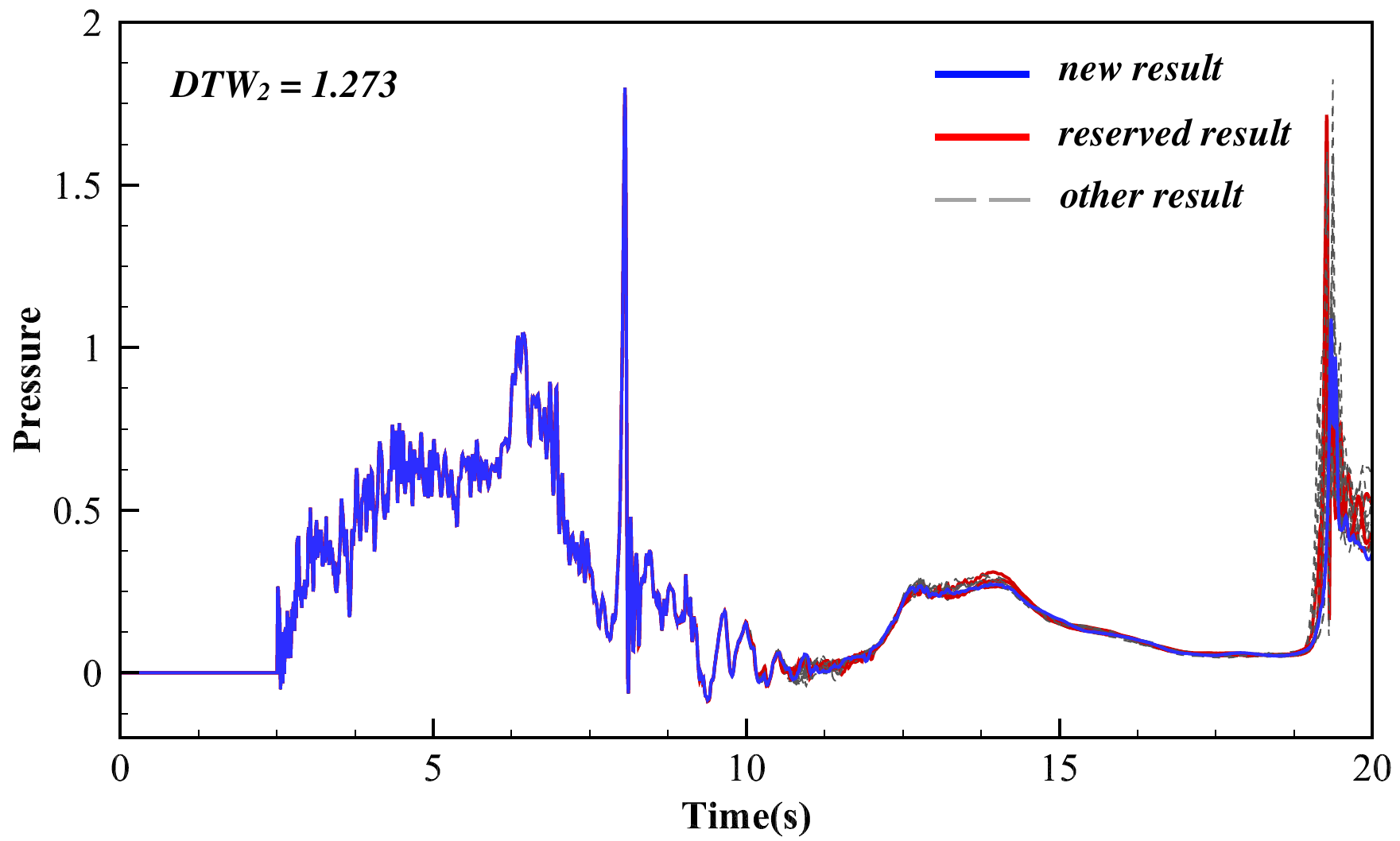}
			\label{pr-re-3db}
		}
		\quad
		\subfigure[]{
			\includegraphics[width=0.5\textwidth]{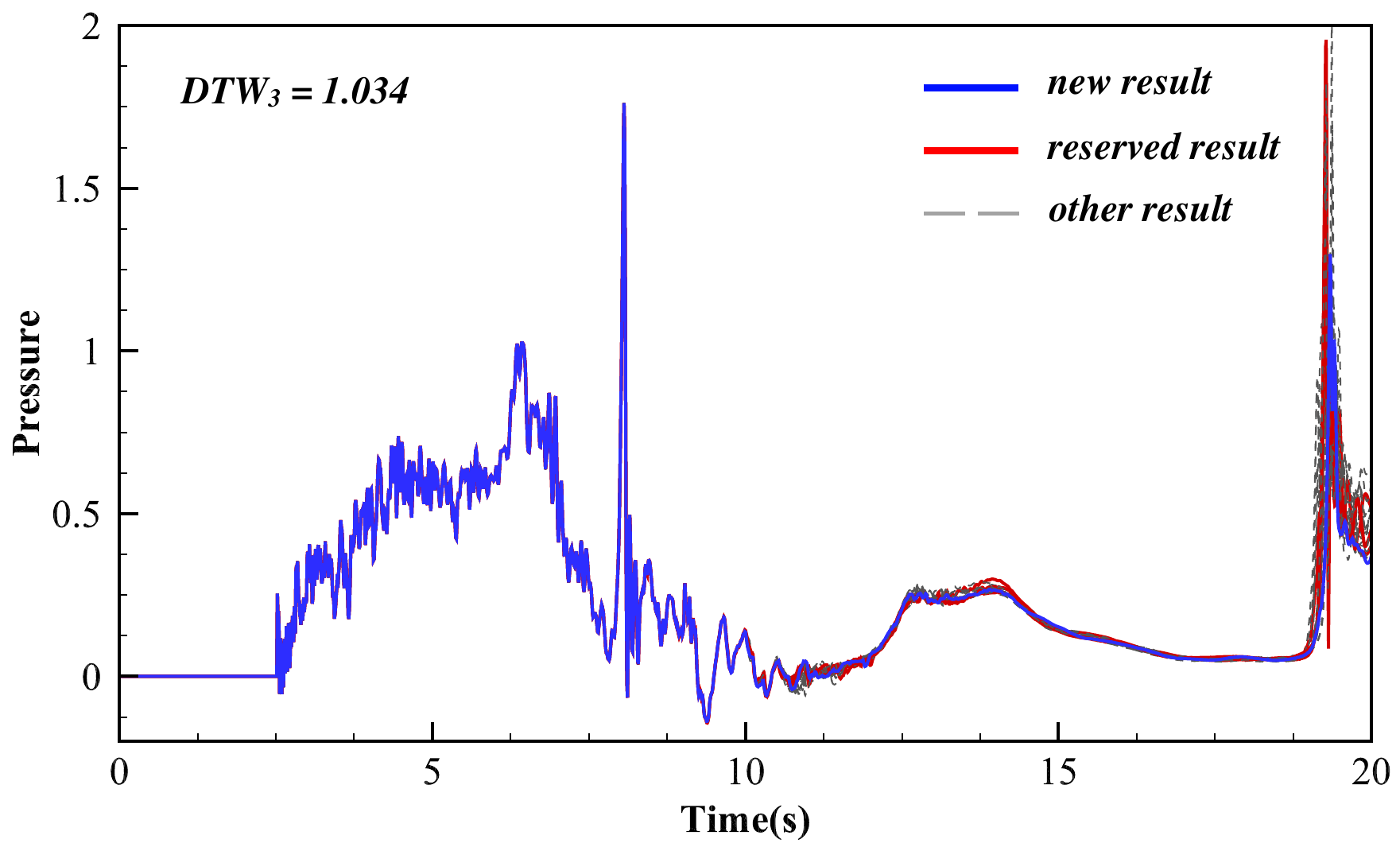}
			\label{pr-re-3dc}
		}
		\quad}
	\caption{Illustration of reference database and results 
		     for the 3d dambreak case: (a) total kinetic energy; (b) pressure at probe a; (c) pressure at probe b; (d) pressure at probe c.}
	\label{dambreak_test_3d}
\end{figure}
It is observed that the kinetic energies obtained after code modification, 
see Figs. \ref{me-re-2d} and \ref{me-re-3d},
are within the range of reference database. 
The collection of multiple pressure results is given in 
Fig. \ref{pr-re-2d} and \ref{pr-re-3da} - \ref{pr-re-3dc},
where the 3d case has three pressure monitoring points. 
After continuously updating the maximum DTW distances for 
each pair of results, the variation of distances converges, 
and the final distance is stored, as listed in Table \ref{dtw_distance}.
Actually, not all computational results but only several randomly chosen 
ones are shown here and have been reserved in the reference database.
\begin{table}[htb!]
	\small
	\renewcommand\arraystretch{1.25}
	\centering
	\captionsetup{font={small}}
	\caption{DTW distance in the reference database and for the new results.}
	\begin{tabularx}{13.5cm}{@{\extracolsep{\fill}}ccccc}
		\hline
		\quad DTW distance & 2d & 3d:probe a & 3d:probe b & 3d:probe c \quad \\
		\midrule
		\quad database & $1.234$ & $0.651$ & $1.273$ & $1.034$ \quad \\
		\quad testing1 & $0.368$ & $0.180$  & $0.229$ & $0.231$ \quad \\
		\quad testing2 & $0.535$ & $0.177$  & $0.153$ & $0.144$ \quad \\
		\quad testing3 & $0.522$ & $0.015$  & $0.084$ & $0.111$ \quad \\
		\bottomrule
	\end{tabularx}
	\label{dtw_distance}
\end{table}
Table \ref{dtw_distance} indicates that the distance between 
the newly obtained results and the ones stored in the database 
are all smaller than the reference distances.
Therefore, after performing the regression test on those two variables, 
new results obtained after code modification are deemed correct,
and the new code is considered to be compatible with the old version 
for the dambreak flow case.
\subsection{Oscillating-beam}\label{subsubsec:beam}
\begin{figure}[htb!]
	\centering
	\makebox[\textwidth][c]{\subfigure[]{
			\includegraphics[width=0.5\textwidth]{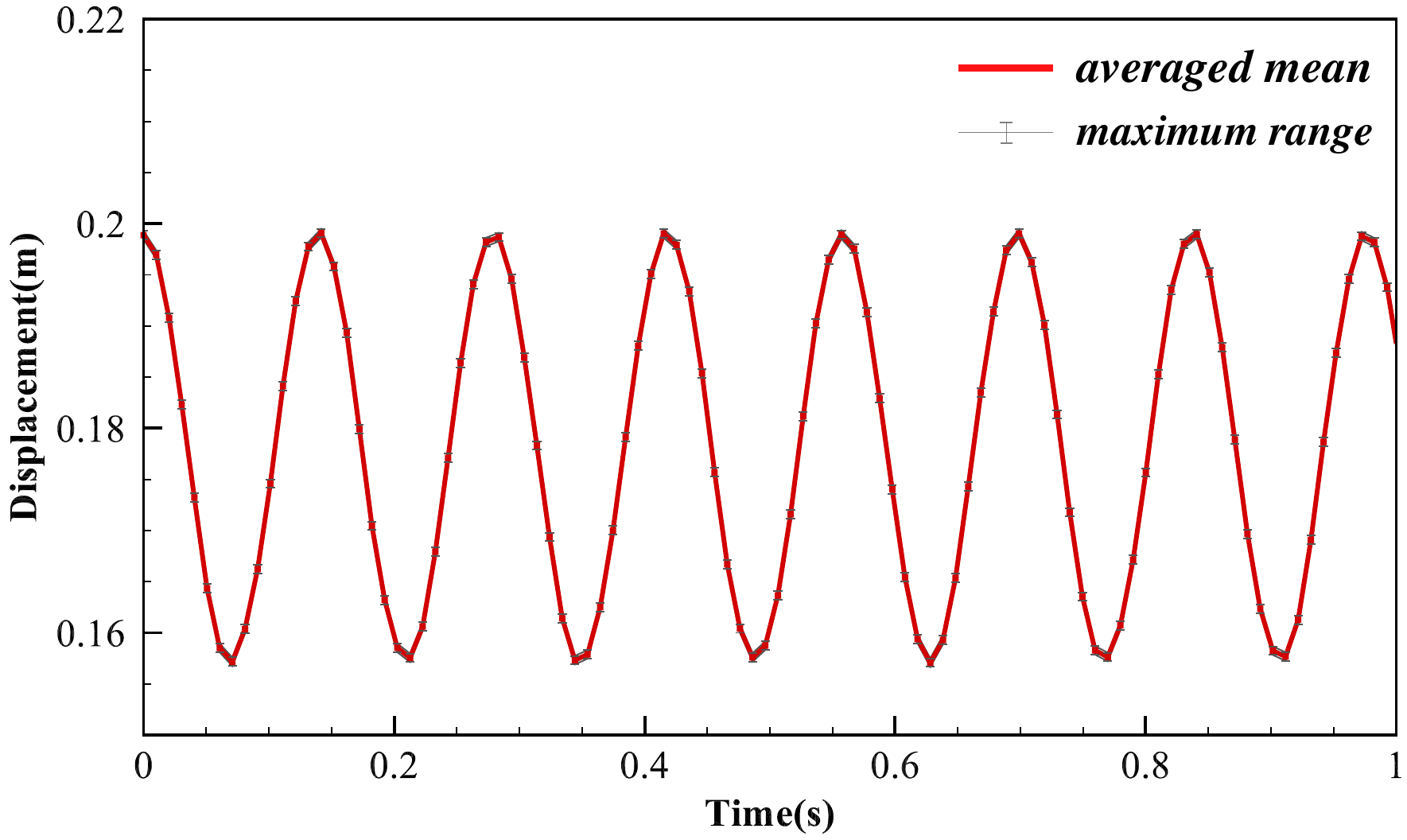}
			\label{position_x}
		}
		\quad
		\subfigure[]{
			\includegraphics[width=0.5\textwidth]{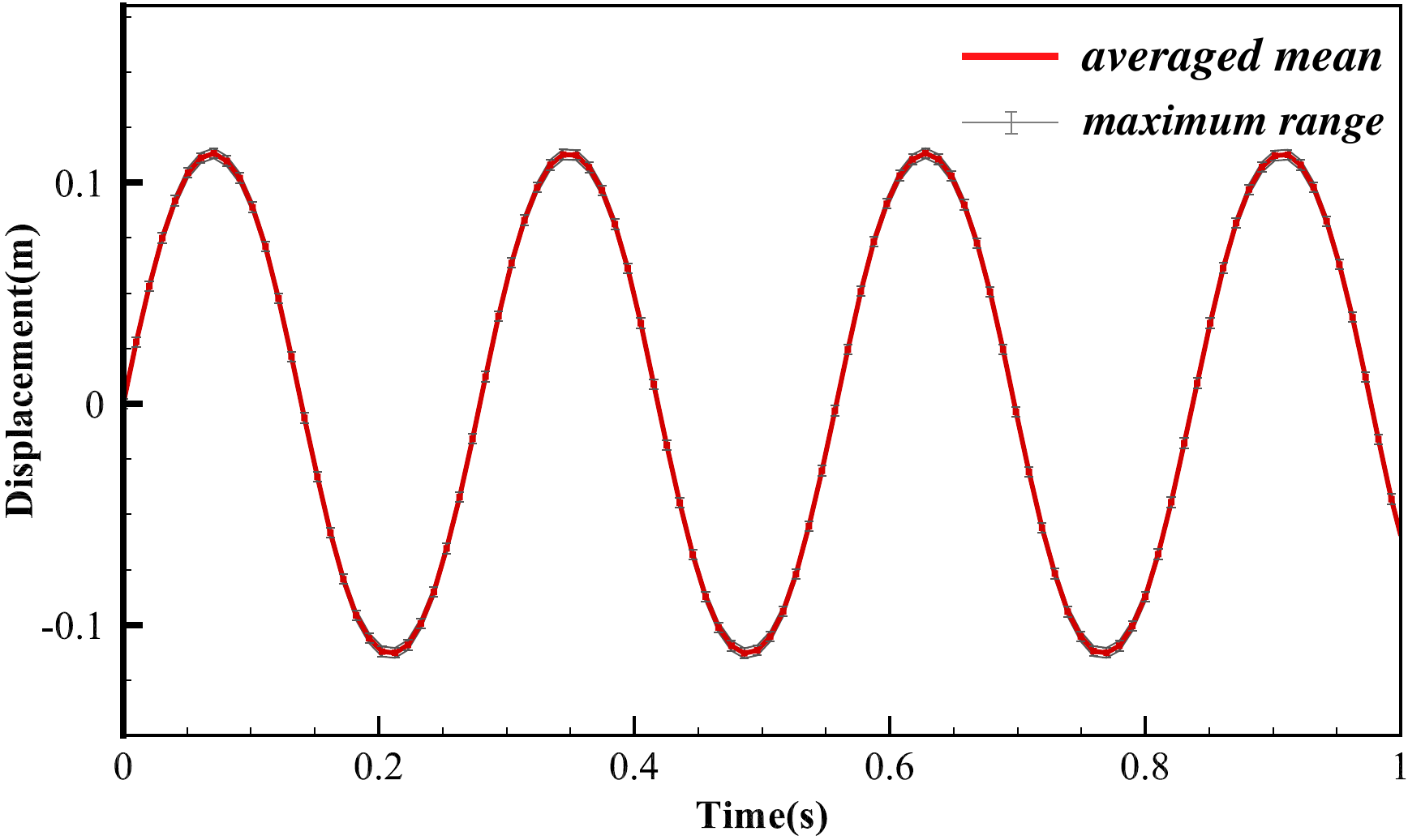}
			\label{position_y}
		}
		\quad}
	\caption{Illustration of the reference database and results for the 2d oscillating beam 
		     case: displacement along (a) $x$ direction, and (b) $y$ direction, respectively. 
		     The red lines represent the ensemble-averaged mean for each data point, 
		     and the gray ones gives the range of maximum variation.}
	\label{oscillating_beam}
\end{figure}

The second example is the free end oscillating elastic beam problem.
The detailed setting up and validation can be referred in our previous work \cite{zhang2017generalized}.
The displacement of the beam tip has been recorded and used for the regression test.
This variable has two components representing different directions,
and it has small differences for each computation,
so the ensemble-averaged method is adopted for generating the 
reference database and new result testing.
Fig. \ref{oscillating_beam} demonstrates the reference database for this case.
It is found that the new result of this case (not shown here due to very small, 
not noticeable visually, differences) after code modifications lies within 
the range given by the reference database for each data point.
\subsection{Fluid-solid interaction}\label{subsubsec:fsi}
\begin{figure}[htb!]
	\centering
	\includegraphics[width=0.95\textwidth]{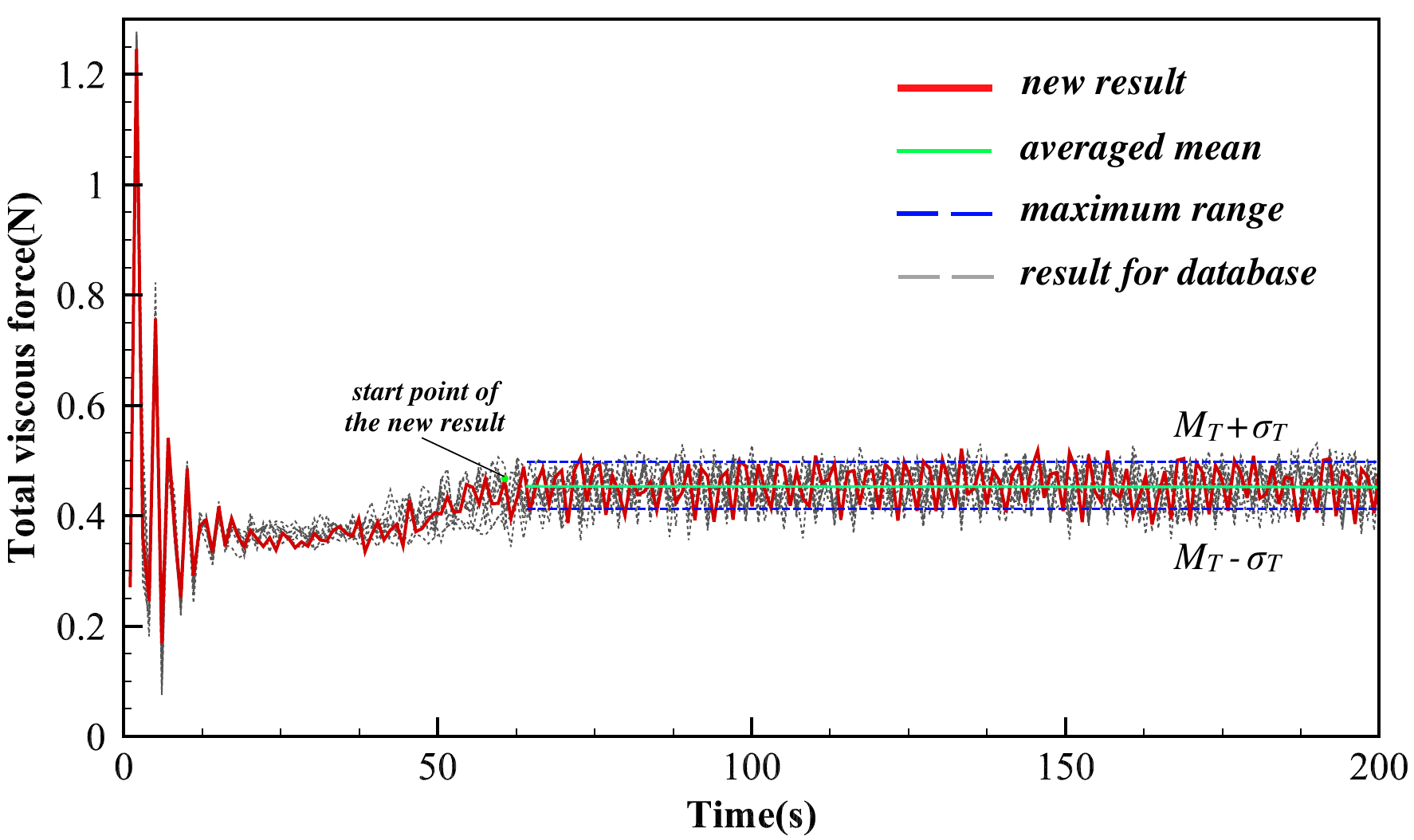}
	\caption{Illustration of the reference database and the tested result for the FSI problem.
		     The gray dashed lines represent multiple results used for generating the database, and the red line is a new result after code modification.
		     The green line shows the time-averaged mean of multiple results,
		     and blue ones indicate the variation range constrained by the variance.}
	\label{fsi_test}
\end{figure}

The last example is a fluid-solid interaction problem on flow-induced vibration.
More information as well as validations can be referred 
in earlier work \cite{zhang2020dual, zhang2021multi}.
The total viscous force from the fluid acting on the solid structure was recorded,
and it fluctuates around the constant value 
when the dynamics entered a periodic oscillation state.
Thus, the time-averaged method is used to perform 
the regression test for this variable.
Since the force in the $y$ direction is relatively small, 
only the $x$ direction force was considered.
Fig. \ref{fsi_test} displays the reference database and 
one tested new result for this case.
Table \ref{time-averaged-test} shows the converged metric values in the database 
as well as the ones from the new result. 
\begin{table}[htb!]
	\small
	\renewcommand\arraystretch{1.25}
	\centering
	\captionsetup{font={small}}
	\caption{Mean and variance in the reference database and for the new result.}
	\begin{tabularx}{10cm}{@{\extracolsep{\fill}}ccc}
		\hline
		\quad metric & database & new result \quad \\
		\midrule
		\quad mean & $0.4557$ & $0.4549$ \quad\\
		\quad variance & $0.00173$ & $0.00157$ \quad\\
		\bottomrule
	\end{tabularx}
	\label{time-averaged-test}
\end{table}
It indicates the new mean is quite close to the converged one, 
and the new variance is also smaller than the reference one.
Therefore, the new result of the FSI problem after code modification 
is still considered correct.

In general, each test case should have at least one variable of interest 
used for the regression test, and the testing strategy is not fixed.
This regression test environment provides flexible combinations of 
variables and strategies, but each variable should always have 
the best option to check itself. 
%
%
\section{Conclusion}\label{conclusion}
This paper introduces a method for developing an automatic 
regression test environment for open-source scientific libraries and 
uses SPHinXsys as an illustration to demonstrate its functionality.
For scientific libraries under centralized development, 
it's essential to guarantee the accuracy of simulation results 
all the time, and the regression test provides this procedure.
The reference database for each benchmark test is generated 
using different strategies, and the new result after code modifications 
can be automatically tested with them once the source code is updated. 
This regression test environment has been implemented in all test cases 
released in SPHinXsys, and it shows great functionality to check the 
validity of the new result obtained after code modifications. 
By doing such work, we also want to drawn some attention from general scientific 
computing communities to emphasize the software performance during development.
The principle of the regression test proposed here is universal 
and can be applied and extended in other libraries and applications. 
In the future, other regression test methods will be implemented, 
and with the number of test cases swelling due to adding new dynamics features, 
selection and reduction of test cases will also be adopted.

\section*{Acknowledgments}
\addcontentsline{toc}{section}{Acknowledgement}
The first author would like to acknowledge the financial support provided by
the China Scholarship Council (No.202006230071).
C. Zhang and X.Y. Hu would like to express their gratitude to Deutsche
Forschungsgemeinschaft(DFG) for their sponsorship of this research under
grant number DFG HU1527/12-4.
The corresponding code of this work is available on GitHub at 
\url{https://github.com/Xiangyu-Hu/SPHinXsys}.

\bibliographystyle{elsarticle-num}
\bibliography{regression}

\end{document}